\documentclass[aps,pra,onecolumn,superscriptaddress,notitlepage,nofootinbib,longbibliography]{revtex4-2}

\usepackage{amsmath,amssymb,amsfonts,bm,mathtools}
\usepackage{graphicx}
\usepackage{float}
\usepackage{psfrag}
\usepackage{mathptmx}
\usepackage{subcaption}
\usepackage{xcolor}
\usepackage[utf8]{inputenc}
\usepackage[T1]{fontenc}
\usepackage{braket}
\usepackage{siunitx}
\usepackage{booktabs}
\usepackage[titletoc,title]{appendix}
\usepackage[shortlabels]{enumitem}
\usepackage[colorlinks=true,linkcolor=blue,citecolor=blue,breaklinks=true]{hyperref}
\usepackage[capitalize,noabbrev]{cleveref}

\begin{document}

\title{Enhanced Magnon Synchronization in Coupled WGM Optomagnonic Resonators with Phase-Dependent Photon Hopping}

\author{Le-Ji Xue}
\affiliation{School of Physical Science and Technology, Nantong University, Nantong 226019, People’s Republic of China}	
\author{Ying-Jian Zhu}
\affiliation{Department of Physics, School of Science, Beihua University, 15, Jilin Str. Jilin, 132013, China}
\author{Jaspal Singh}
\affiliation{Laboratory for Advanced Nanostructures for Environmental and Energy Applications, Department of Physics, Akal University, Talwandi Sabo, Raman Road, Bathinda, 151302, Punjab, INDIA}
\author{Ahmad Zahia}
\affiliation{Department of Mathematics, Faculty of Science, Benha University, Benha, Egypt}
\author{Kong-Ming Hu}
\affiliation{School of Physical Science and Technology, Nantong University, Nantong 226019, People’s Republic of China}
\author{Jia-Xin Peng}
\affiliation{School of Physical Science and Technology, Nantong University, Nantong 226019, People’s Republic of China}
\author{S. K. Singh}\email{singhshailendra3@gmail.com}\thanks{(Corresponding Author)}
\affiliation{Department of Physics, Akal University, Talwandi Sabo, Punjab, India}

\date{\today}

\begin{abstract}

We investigate quantum synchronization in a coupled cavity optomagnonic system which consists of two spatially separated optical whispering-gallery-mode (WGM) resonators and each  resonator is also coupled to a yttrium iron garnet (YIG) sphere through the optomagnonic interaction. 
Phase-dependent single-photon hopping factor couples the two optical resonators and provides an indirect interaction between the two distant magnon modes. We then investigate complete synchronization, $\phi$-synchronization, and quantum phase synchronization using the covariance-matrix formalism as well as also studied the effects of hopping term on the overall synchronization dynamics of two distant magnon modes. It can be seen that  the photon-hopping phase provides an efficient way  to control the synchronization dynamics and when it is varied from $0$ to $\pi$, the magnon trajectories gradually evolve from weakly correlated motion to a highly synchronized state which is also accompanied by a significant reduction in the synchronization error. The influence of the photon-hopping strength and thermal fluctuations is also investigated where it can be seen that  stronger photon hopping enhances all synchronization measures whereas thermal noise weakens the coherent correlations responsible for synchronized dynamics. Our results demonstrate that the phase of hopping factor offers a simple and effective approach for controlling synchronization dynamics in WGM based coupled cavity optomagnonic systems and also  provide a useful route toward coherent control of collective magnon dynamics in such quantum optomganonic devices.

\end{abstract}

\maketitle

\section{Introduction}

Synchronization is a fundamental collective phenomenon in which interacting systems adjust their rhythms and evolve in a correlated manner. Since the pioneering observation of synchronization between two pendulum clocks by Huygens \cite{1}, the subject has attracted considerable attention in a wide range of physical systems. Extending synchronization concepts to the quantum regime is particularly challenging due to the constraints imposed by the Heisenberg uncertainty principle. A major advance in this direction was achieved by Mari \textit{et al.}, who introduced quantitative measures of complete quantum synchronization and quantum phase synchronization for continuous-variable systems \cite{2,63}. Owing to their intrinsic nonlinear dynamics and self-sustained oscillations, cavity optomechanical systems have emerged as a natural platform for investigating quantum synchronization \cite{3,4}. Since then, synchronization phenomena have been explored in several quantum platforms, including cavity quantum electrodynamics systems, atomic ensembles, van der Pol oscillators, Bose-Einstein condensates, and superconducting circuits \cite{5,7,10,13,14}. Beyond its fundamental significance, quantum synchronization has found potential applications in quantum communication, sensing, and information processing \cite{16,19,22}. Furthermore, its close connection with quantum correlations such as mutual information, quantum discord, and entanglement has stimulated extensive research efforts in recent years \cite{6,24,26a,28a,30a} whereas the various physical platforms which are available for realizing quantum synchronization phenomena, it can be said that the hybrid cavity optomagnonic systems have recently emerged as particularly attractive platforms due to their strong light-matter interaction mechanism, long coherence times, and high degree of tunability.

Recently, cavity optomagnonic systems have attracted increasing attention as a versatile platform for exploring macroscopic quantum phenomena. In particular, ferrimagnetic materials such as YIG sphere possess excellent magnetic properties together with exceptionally low dissipation rates which make them well suited for coherent light-matter interactions \cite{26,27}. We would like to mention here that collective spin excitations in YIG, known as magnons, possess frequencies that can be conveniently tuned by external magnetic fields, making them highly versatile for quantum control and information processing applications. The large number of participating spins enables strong coupling between magnons and electromagnetic fields, facilitating the realization of strong-coupling regimes in cavity magnonic systems \cite{26,27,28,29}. In particular, the Kittel mode of a YIG sphere exhibits long coherence times and low damping rates, which have motivated extensive investigations of cavity magnonics and magnomechanics \cite{30,31,32}. As a consequence, a variety of fascinating quantum phenomena have been reported, including four-wave mixing \cite{33}, magnomechanically induced transparency and absorption \cite{28,34}, bipartite and multipartite entanglement generation \cite{36,41,42}, Einstein-Podolsky-Rosen entanglement and quantum steering \cite{46}, as well as nonlinear effects arising from magnon Kerr interactions \cite{44,45}. More recently, the successful realization of strong magnon-photon coupling in both microwave and optical domains has opened new directions for studying hybrid quantum systems \cite{26,27,29,47,48,49,50,51,52,53,54,55}. This is very important to mention here that unlike microwave cavity based platforms, where the interaction is typically described by a beam-splitter-type coupling, cavity optomagnonic systems based on optical WGM resonators exhibit an interaction analogous to the radiation-pressure coupling encountered in cavity optomechanics \cite{47,48,49}. In these systems, the optical field interacts with collective spin excitations through magneto-optical effects, and the coupling strength can be further enhanced using pump--probe techniques \cite{51}. Consequently, cavity optomagnonic platforms have become promising candidates for investigating a wide range of nonclassical phenomena, including optical squeezing, entanglement generation, quantum steering, and coherent light-magnon interaction \cite{47,48,49,50,56}. These seminal and remarkable theoretical and experimental advances demonstrate that cavity optomagnonic systems provide a versatile platform for exploring a variety of macroscopic quantum phenomena which is because of their high controllability, long coherence times, and strong photon-magnon interaction.\\
In this work, we consider two spatially separated optical WGM resonators where each one contains a YIG sphere as well as  two optical cavities are coupled through phase-dependent single-photon hopping. Such a configuration allows the two magnon modes to interact indirectly through the optical fields. We investigate complete synchronization, $\phi$-synchronization, quantum phase synchronization, and also explore the effects of phase dependent photon hopping on the overall synchronization behaviour of the coupled WGM resonators.
The remainder of this paper is organized as follows. In Sec.~II, we introduce the theoretical model and derive the corresponding quantum Langevin equations. Section~III presents the synchronization measures and the covariance-matrix formalism employed to study synchronisation phenomena. We give the numerical results and detailed discussion  in Sec.~IV which is also followed by the experimental feasibility of the proposed scheme in Sec.~V. Finally, the main conclusions are summarized and given in Sec.~VI. 

\section{Model Hamiltonian and Quantum Dynamics}

\begin{figure}[t]
\centering
\includegraphics[width=\linewidth]{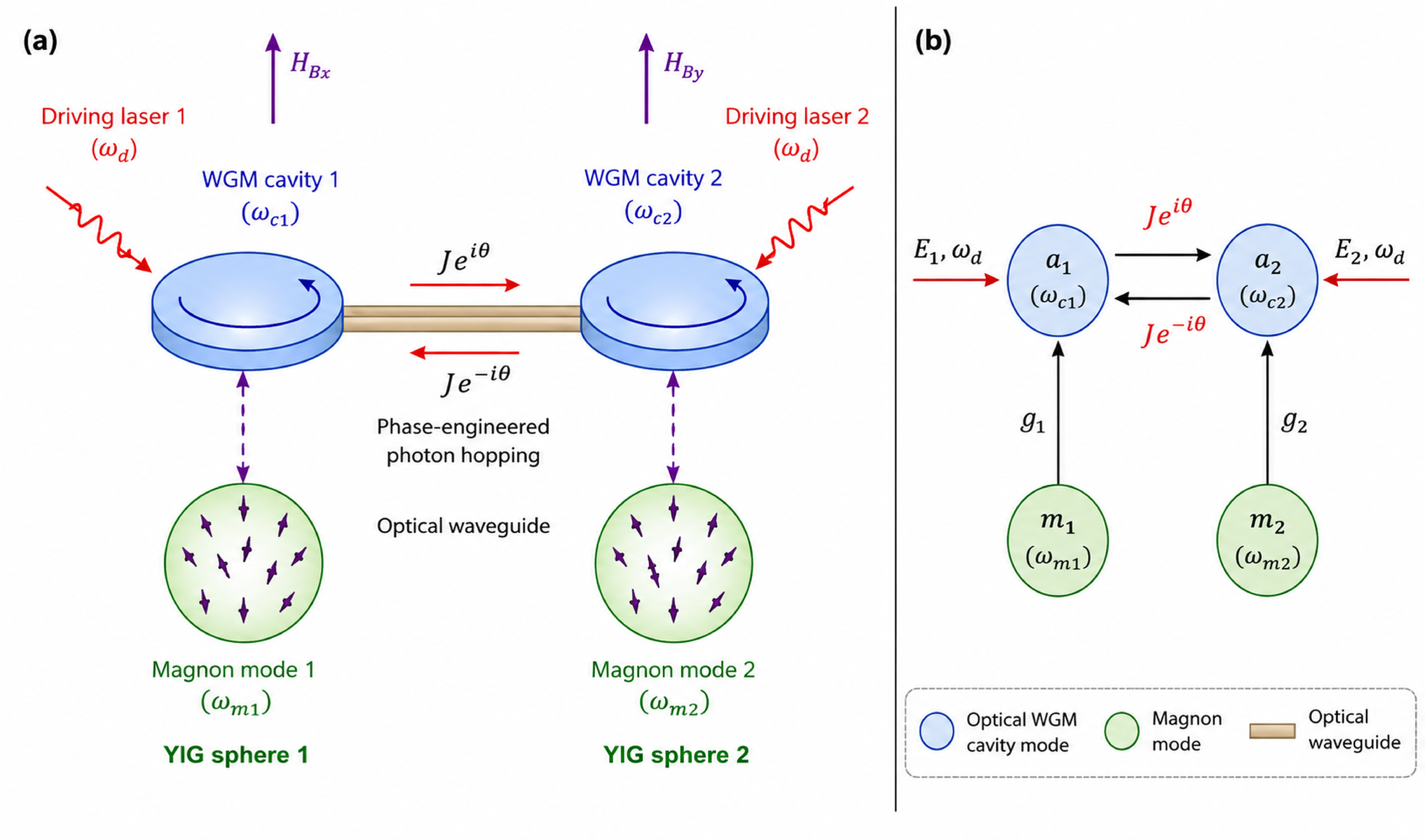}
\caption{
(a) Schematic illustration of the coupled cavity optomagnonic system consisting of two WGM cavities interacting with two magnon modes hosted in YIG resonators and coupled through phase-dependent photon hopping. (b) Simplified coupling diagram showing the optical cavity modes, magnon modes, and the phase-engineered photon-hopping interaction $Je^{\pm i\theta}$.
}
\label{fig1}
\end{figure}

As illustrated in  Fig.\ref{fig1}, this proposed system consists of two spatially separated WGM optomagnonic resonators coupled through an optical waveguide that enables coherent photon transport between the cavities. Each resonator supports a high-quality optical WGM interacting with a localized magnon mode hosted by a  YIG sphere \cite{26,27,57}. The magnon modes are excited by external bias magnetic fields, denoted by $H_{B_{x1}}$ and $H_{B_{x2}}$, respectively. The interaction between the optical and magnonic degrees of freedom originates from the magneto-optical Faraday effect, which couples the collective spin excitations of the magnetic medium to the circulating optical field within the WGM resonators \cite{47,48,49,50,51,56,57}. In optical WGM-based cavity optomagnonic systems, the interaction between the collective spins associated with the magnon modes and the circularly polarized optical WGM fields can be described by the interaction Hamiltonians $H_{B_{x1}}=\hbar a_1^\dagger a_1 G_{x1}S_{x1}$ and $H_{B_{x2}}=\hbar a_2^\dagger a_2 G_{x2}S_{x2}$ \cite{50,56,57}, where $a_j$ ($a_j^\dagger$) denotes the annihilation (creation) operator of the optical mode, $G_{xj}$ represents the magneto-optical coupling coefficient, and $S_{xj}$ corresponds to the collective spin operator of the $j$th magnon mode. The optical WGM field couples to the transverse spin component of the magnetization, giving rise to the Faraday interaction. The corresponding coupling strength can be expressed as $G_j=c\Theta_{f_j}/(4S_j\epsilon_{r_j})$, where $\Theta_{f_j}$ denotes the Faraday rotation per unit length, $c$ is the speed of light in vacuum, $S_j$ is the total spin of the YIG sphere, and $\epsilon_{r_j}$ is the relative permittivity of the magnetic material \cite{56,57}.\\

In the low-excitation regime, where the average number of excited magnons remains much smaller than the total spin population of the YIG sphere, i.e., $m_j^\dagger m_j \ll S_j$, the Holstein-Primakoff transformation can be easily employed to map the collective spin operators onto bosonic magnon operators \cite{60}. Under this approximation, the spin operators can be written as $S_{xj}\simeq \sqrt{S_j/2},(m_j+m_j^\dagger)$, where $m_j$ ($m_j^\dagger$) denotes the annihilation (creation) operator of the magnon mode whereas the magnon mode of interest corresponds to the homogeneous Kittel mode \cite{30,31}. In addition to the local optomagnonic interaction, the optical cavity modes are coupled through a waveguide-assisted photon-hopping channel. The photon hopping carries a controllable phase factor $\theta$, which can be adjusted by changing the optical path or by using suitable phase-control techniques. This phase provides an additional way to control the transport of photons between the two cavities and hence the synchronization dynamics between the two distant magnon modes.

In a frame rotating at the driving laser frequency $\omega_d$, the total Hamiltonian of the system can be written as
\begin{equation}
H = H_0 + H_{\rm int} + H_J + H_d,
\end{equation}
where
\begin{align}
H_0 &=
-\hbar \Delta_1 a_1^\dagger a_1
-\hbar \Delta_2 a_2^\dagger a_2
+\hbar \omega_{m1} m_1^\dagger m_1
+\hbar \omega_{m2} m_2^\dagger m_2, \nonumber\\
H_{\rm int}
&=
\hbar g_1 a_1^\dagger a_1(m_1+m_1^\dagger)
+
\hbar g_2 a_2^\dagger a_2(m_2+m_2^\dagger), \nonumber\\
H_J
&=
\hbar J
\left(
e^{i\theta}a_1^\dagger a_2
+
e^{-i\theta}a_2^\dagger a_1
\right), \nonumber\\
H_d
&=
i\hbar E_1(a_1^\dagger-a_1)
+
i\hbar E_2(a_2^\dagger-a_2).
\end{align}

Here, $a_j$ ($a_j^\dagger$) and $m_j$ ($m_j^\dagger$) denote the annihilation (creation) operators of the optical cavity and magnon modes with frequencies $\omega_{cj}$ and $\omega_{mj}$, respectively. The optical and magnon modes obey the standard bosonic commutation relations $[o,o^\dagger]=1$, where $o=a,m$. The cavity-laser detuning is defined as $\Delta_j=\omega_d-\omega_{cj}$, with $\omega_d$ denoting the frequency of the external driving laser. Furthermore, $\omega_{mj}$ is determined by the corresponding bias magnetic field through the relation $\omega_{mj}=\gamma H_{B_{xj}}$, where $\gamma$ is the gyromagnetic ratio. The parameter $g_j$ represents the single-photon optomagnonic coupling strength between the optical WGM cavity and the $j$th magnon mode \cite{47,48,49,50,56,57} whereas $J$ denotes the photon-hopping term and $\theta$ its phase. Finally, the last term describes the coherent driving of the optical cavities by external laser fields with amplitudes $E_1$ and $E_2$, which are related to the input laser powers through $E_j=\sqrt{2\kappa_j P_j/\hbar\omega_d}$, where $P_j$ denotes the input laser power and $\kappa_j$ is the optical cavity decay rate.

We employ the well-known quantum Langevin equations (QLEs) to investigate the quantum dynamics of the proposed system while accounting for the dissipation and the noise associated with the optical cavity and magnon modes  \cite{61}. The cavity decay rates $\kappa_j$ and magnon damping rates $\gamma_j$ are included in the dynamical description whereas the corresponding Heisenberg--Langevin equations can therefore be written as

\begin{eqnarray}
\dot a_1
&=&
(i\Delta_1-\kappa_1)a_1
-i g_1 a_1(m_1+m_1^\dagger)
-iJ e^{i\theta}a_2
+E_1
+\sqrt{2\kappa_1}\,a_1^{\rm in},
\nonumber\\
\dot a_2
&=&
(i\Delta_2-\kappa_2)a_2
-i g_2 a_2(m_2+m_2^\dagger)
-iJ e^{-i\theta}a_1
+E_2
+\sqrt{2\kappa_2}\,a_2^{\rm in},
\nonumber\\
\dot m_1
&=&
-(i\omega_{m1}+\gamma_1)m_1
-i g_1 a_1^\dagger a_1
+\sqrt{2\gamma_1}\,m_1^{\rm in},
\nonumber\\
\dot m_2
&=&
-(i\omega_{m2}+\gamma_2)m_2
-i g_2 a_2^\dagger a_2
+\sqrt{2\gamma_2}\,m_2^{\rm in}.
\label{eq:QLE}
\end{eqnarray}
Here, $a^{\mathrm{in}}_{j}$ and $m^{\mathrm{in}}_{j}$ denote the input noise operators associated with the $j$th optical cavity and magnon mode, respectively. These operators possess zero mean values, while their non-vanishing correlation functions are given by  \cite{61}

\begin{eqnarray}
\langle a_{j}^{\mathrm{in}}(t)a_{j}^{\mathrm{in},\dagger}(t') \rangle
&=& \delta(t-t'), \nonumber\\
\langle m_{j}^{\mathrm{in},\dagger}(t)m_{j}^{\mathrm{in}}(t') \rangle
&=& n_{\mathrm{th},j}\delta(t-t'), \nonumber\\
\langle m_{j}^{\mathrm{in}}(t)m_{j}^{\mathrm{in},\dagger}(t') \rangle
&=& \left(n_{\mathrm{th},j}+1\right)\delta(t-t').
\label{eq:noise_correlations}
\end{eqnarray}

Here, $\left[\exp\left(\frac{\hbar\omega_{mj}}{k_B T}\right)-1\right]^{-1} $ denotes the mean thermal occupation number of the $j$th magnon mode, where $\omega_{mj}$ is the corresponding magnon frequency, $T$ is the environmental temperature, and $k_B$ is the Boltzmann constant \cite{61}.

Now we introduce the dimensionless magnon quadratures $q_j=(m_j+m_j^\dagger)/\sqrt{2}$ and $p_j=(m_j-m_j^\dagger)/(i\sqrt{2})$, together with the optical quadratures $X_j=(a_j+a_j^\dagger)/\sqrt{2}$ and $Y_j=(a_j-a_j^\dagger)/(i\sqrt{2})$ \cite{61}. In terms of these quadrature operators, the quantum Langevin equations can be expressed in an easy analytical form which is  suitable for analysing the synchronization dynamics of the indirectly coupled magnon modes. The optomagnonic interaction give rise to the effective cavity detunings $\Delta_1^{\rm eff}=\Delta_{1}-\sqrt{2}g_{1}q_{1}$ and $\Delta_{2}^{\rm eff}=\Delta_{2}-\sqrt{2}g_{2}q_{2}$, and the corresponding equations of motion can be written as

\begin{eqnarray}
\dot q_1
&=&
\omega_{m1}p_1,
\nonumber\\
\dot p_1
&=&
-\omega_{m1}q_1
+\sqrt{2}g_1 n_1
-\gamma_1p_1
+\sqrt{2\gamma_1}\,p_{1}^{\rm in},
\nonumber\\
\dot X_1
&=&
-\kappa_1X_1
+\Delta_1^{\rm eff}Y_1
+
J\left(\sin\theta\,X_2+\cos\theta\,Y_2\right)
+
E_1
+\sqrt{2\kappa_1}\,X_{1}^{\rm in},
\nonumber\\
\dot Y_1
&=&
-\kappa_1Y_1
-\Delta_1^{\rm eff}X_1
+
J\left(\sin\theta\,Y_2-\cos\theta\,X_2\right)
+\sqrt{2\kappa_1}\,Y_{1}^{\rm in},
\nonumber\\
\dot q_2
&=&
\omega_{m2}p_2,
\nonumber\\
\dot p_2
&=&
-\omega_{m2}q_2
+\sqrt{2}g_2 n_2
-\gamma_2p_2
+\sqrt{2\gamma_2}\,p_{2}^{\rm in},
\nonumber\\
\dot X_2
&=&
-\kappa_2X_2
+\Delta_2^{\rm eff}Y_2
+
J\left(-\sin\theta\,X_1+\cos\theta\,Y_1\right)
+
E_2
+\sqrt{2\kappa_2}\,X_{2}^{\rm in},
\nonumber\\
\dot Y_2
&=&
-\kappa_2Y_2
-\Delta_2^{\rm eff}X_2
+
J\left(-\sin\theta\,Y_1-\cos\theta\,X_1\right)
+\sqrt{2\kappa_2}\,Y_{2}^{\rm in}.
\end{eqnarray}
where $n_j=a_j^\dagger a_j = \frac{X_j^2+Y_j^2}{2}$ denotes the intracavity photon number of the $j$th optical cavity mode. The corresponding semiclassical steady-state values are obtained from the classical dynamical equations by replacing the operators with their mean values and neglecting the associated noise terms. The resulting equations are given as \cite{57}
\begin{eqnarray}
\dot q_{1s}
&=&
\omega_{m1}p_{1s},
\nonumber\\
\dot p_{1s}
&=&
-\omega_{m1}q_{1s}
+\sqrt{2}g_1n_{1s}
-\gamma_1p_{1s},
\nonumber\\
\dot X_{1s}
&=&
-\kappa_1X_{1s}
+\Delta_{1}^{\rm eff}Y_{1s}
+
J(\sin\theta\,X_{2s}+\cos\theta\,Y_{2s})
+
E_1,
\nonumber\\
\dot Y_{1s}
&=&
-\kappa_1Y_{1s}
-\Delta_{1}^{\rm eff}X_{1s}
+
J(\sin\theta\,Y_{2s}-\cos\theta\,X_{2s}),
\nonumber\\
\dot q_{2s}
&=&
\omega_{m2}p_{2s},
\nonumber\\
\dot p_{2s}
&=&
-\omega_{m2}q_{2s}
+\sqrt{2}g_2n_{2s}
-\gamma_2p_{2s},
\nonumber\\
\dot X_{2s}
&=&
-\kappa_2X_{2s}
+\Delta_{2}^{\rm eff}Y_{2s}
+
J(-\sin\theta\,X_{1s}+\cos\theta\,Y_{1s})
+
E_2,
\nonumber\\
\dot Y_{2s}
&=&
-\kappa_2Y_{2s}
-\Delta_{2}^{\rm eff}X_{2s}
+
J(-\sin\theta\,Y_{1s}-\cos\theta\,X_{1s}),
\end{eqnarray}

The steady-state solutions are obtained by setting
$\dot q_{js}=\dot p_{js}=\dot X_{js}=\dot Y_{js}=0$.
These steady-state values are then used to linearise the nonlinear quantum Langevin equations around the corresponding operating point.

To investigate the quantum synchronization properties of the system, we linearise the dynamics around the semi-classical steady state where each dynamical variable is decomposed into its steady-state value and a small quantum fluctuation according to  \cite{61}
\begin{equation}
u=u_s+\delta u,
\end{equation}
where
\begin{equation}
u=
(q_1,p_1,X_1,Y_1,q_2,p_2,X_2,Y_2)^T.
\end{equation}
Substituting these expressions into Eq.~(5) and retaining only the first-order fluctuation terms yields the linearized Langevin equation
\begin{equation}
\dot{\delta u}=A\,\delta u+n(t),
\end{equation}
where $A$ is the corresponding drift matrix and
\begin{equation}
n(t)=
\Big(
0,\,
\sqrt{2\gamma_1}\,p_1^{\rm in},\,
\sqrt{2\kappa_1}\,X_1^{\rm in},\,
\sqrt{2\kappa_1}\,Y_1^{\rm in},\,
0,\,
\sqrt{2\gamma_2}\,p_2^{\rm in},\,
\sqrt{2\kappa_2}\,X_2^{\rm in},\,
\sqrt{2\kappa_2}\,Y_2^{\rm in}
\Big)^T
\end{equation}
denotes the respective noise vector.

To simplify the notation, we introduce the effective optomagnonic coupling parameters $G_{1r}=g_1X_{1s}$, $G_{1i}=g_1Y_{1s}$, $G_{2r}=g_2X_{2s}$, and $G_{2i}=g_2Y_{2s}$, where $X_{js}$ and $Y_{js}$ denote the steady-state values of the optical quadratures. In addition, we use the shorthand notations $J_{c}=J\cos\theta$ and $J_{s}=J\sin\theta$ for the phase-dependent photon-hopping terms. With these definitions, the drift matrix governing the linearised dynamics can be written as 

\begin{equation}
A=
\begin{pmatrix}
0 & \omega_{m1} & 0 & 0 & 0 & 0 & 0 & 0\\
-\omega_{m1} & -\gamma_1 &
\sqrt2G_{1r} &
\sqrt2G_{1i} &
0 & 0 & 0 & 0\\
-\sqrt2G_{1i} & 0 &
-\kappa_1 &
\Delta_1^{\rm eff} &
0 & 0 &
J_{s} & J_{c}\\
\sqrt2G_{1r} & 0 &
-\Delta_1^{\rm eff} &
-\kappa_1 &
0 & 0 &
-J_{c} & J_{s}\\
0 & 0 & 0 & 0 &
0 & \omega_{m2} &
0 & 0\\
0 & 0 & 0 & 0 &
-\omega_{m2} &
-\gamma_2 &
\sqrt2G_{2r} &
\sqrt2G_{2i}\\
0 & 0 &
-J_{s} &
J_{c} &
-\sqrt2G_{2i} &
0 &
-\kappa_2 &
\Delta_2^{\rm eff}\\
0 & 0 &
-J_{c} &
-J_{s} &
\sqrt2G_{2r} &
0 &
-\Delta_2^{\rm eff} &
-\kappa_2
\end{pmatrix}.
\end{equation}

\section{Covariance Matrix Dynamics and Synchronization Measures}

The linearised dynamics of the fluctuation operators is governed by Eq.~(8). Following the standard treatment of linear quantum Langevin equations as given in \cite{61}  we can write the formal solution can be written as

\begin{equation}
\delta u(t)
=
M(t)\delta u(0)
+
\int_{0}^{t}
M(t-s)\,
n(s)\,
ds,
\label{eq:formal_solution}
\end{equation}

where

\begin{equation}
M(t)=e^{At}
\end{equation}

is the propagator associated with the drift matrix $A$.

The Gaussian quantum fluctuations of the coupled optomagnonic system are completely characterized by the covariance matrix \cite{61}

\begin{equation}
V_{ij}(t)
=
\frac{1}{2}
\Big\langle
\delta u_i(t)\delta u_j(t)
+
\delta u_j(t)\delta u_i(t)
\Big\rangle ,
\label{eq:covariance}
\end{equation}

where $\delta u_i$ denotes the $i$th component of the fluctuation vector.

Assuming Markovian noise correlations \cite{61},

\begin{equation}
\langle
n_i(t)n_j(t')
\rangle
=
D_{ij}\,
\delta(t-t'),
\end{equation}

the covariance matrix evolves according to the Lyapunov equation \cite{61}

\begin{equation}
\dot V(t)
=
A\,V(t)
+
V(t)A^{T}
+
D,
\label{eq:Lyapunov}
\end{equation}

where the diffusion matrix is given as follows \cite{61}

\begin{equation}
D=
{\rm diag}
\Big(
0,
\gamma_1(2\bar n_{\rm th,1}+1),
\kappa_1,
\kappa_1,
0,
\gamma_2(2\bar n_{\rm th,2}+1),
\kappa_2,
\kappa_2
\Big).
\label{eq:diffusion}
\end{equation}

Here,

\begin{equation}
\bar n_{\rm th}
=
\left[
\exp
\left(
\frac{\hbar\omega_m}
{k_B T}
\right)
-1
\right]^{-1}
\end{equation}

denotes the mean thermal magnon occupation number, where $T$ is the environmental temperature and $k_B$ is the Boltzmann constant.

For a stable dynamical regime, the covariance matrix approaches a stationary value in the long-time limit \cite{61},

\begin{equation}
V(t)
\rightarrow
V_{\rm ss},
\qquad
t\rightarrow\infty,
\end{equation}

where the steady-state covariance matrix satisfies \cite{61}

\begin{equation}
A V_{\rm ss}
+
V_{\rm ss}A^{T}
+
D
=
0.
\label{eq:ssCM}
\end{equation}

To quantify synchronization between the two magnon modes, we introduce the collective error quadratures

\begin{equation}
q_-(t)
=
\frac{q_1(t)-q_2(t)}
{\sqrt{2}},
\qquad
p_-(t)
=
\frac{p_1(t)-p_2(t)}
{\sqrt{2}}.
\end{equation}

Following the synchronization measure introduced by Mari \textit{et al.}~\cite{2}, the complete synchronization measure is defined as

\begin{equation}
S_c(t)
=
\left\langle
q_-^2(t)
+
p_-^2(t)
\right\rangle^{-1}.
\label{eq:Sc_definition}
\end{equation}

To characterize $\phi$-synchronization, we introduce the phase-rotated quadratures following Ref.~\cite{63}

\begin{equation}
q_j'
=
q_j\cos\phi_j
+
p_j\sin\phi_j,
\qquad
p_j'
=
p_j\cos\phi_j
-
q_j\sin\phi_j,
\end{equation}

where

\begin{equation}
\phi_j
=
\tan^{-1}
\left(
\frac{p_{js}}
{q_{js}}
\right)
\end{equation}

is the phase associated with the steady-state trajectory of the $j$th magnon mode \cite{63}. The relative phase quadrature is then defined as

\begin{equation}
p_-'
=
\frac{p_1'-p_2'}
{\sqrt{2}},
\end{equation}

which leads to the quantum phase synchronization measure introduced in Ref.~\cite{2}

\begin{equation}
S_p(t)
=
\frac{1}
{2\langle (p_-')^2\rangle}.
\label{eq:Sp_definition}
\end{equation}

Using the covariance-matrix formalism, the synchronization measures can be expressed entirely in terms of the covariance-matrix elements \cite{2,63}. The complete synchronization measure becomes

\begin{equation}
S_c(t)
=
\left[
\frac{1}{2}
\left(
V_{11}
+
V_{22}
+
V_{55}
+
V_{66}
-
2V_{15}
-
2V_{26}
\right)
\right]^{-1},
\label{eq:Sc}
\end{equation}

Following Ref.~\cite{63}, the $\phi$-synchronization measure is given by

\begin{equation}
S_c^{\phi}(t)
=
\left[
\frac{1}{2}
\left(
V_{11}
+
V_{22}
+
V_{55}
+
V_{66}
+
2V_{25}\sin\phi
-
2V_{16}\sin\phi
-
2V_{26}\cos\phi
-
2V_{15}\cos\phi
\right)
\right]^{-1},
\label{eq:Sphi}
\end{equation}

where

\begin{equation}
\phi
=
\phi_1-\phi_2
\end{equation}

is the relative phase between the two magnon oscillators.

Finally, the quantum phase synchronization measure can be written as \cite{2,63}

\begin{equation}
\begin{aligned}
S_p(t)
=
\Big[
&
V_{11}\sin^2\phi_1
+
V_{22}\cos^2\phi_1
+
V_{55}\sin^2\phi_2
+
V_{66}\cos^2\phi_2
\\
&
-2V_{12}\cos\phi_1\sin\phi_1
-2V_{15}\sin\phi_1\sin\phi_2
+2V_{16}\sin\phi_1\cos\phi_2
\\
&
+2V_{25}\cos\phi_1\sin\phi_2
-2V_{26}\cos\phi_1\cos\phi_2
-2V_{56}\cos\phi_2\sin\phi_2
\Big]^{-1}.
\end{aligned}
\label{eq:Sp}
\end{equation}

The quantities $S_c(t)$, $S_c^{\phi}(t)$, and $S_p(t)$ characterize complete synchronization, phase synchronization, and quantum phase synchronization, respectively \cite{2,63}.

\section{Results and Discussion}

In this section, we investigate the dynamical behaviour and synchronization properties of the proposed coupled WGM-based optomagnonic system by numerically solving Eq.~(6) where we have set $\omega_{m1}=1$ which means that all frequencies, damping rates, coupling strengths, photon-hopping amplitudes, and driving strengths are normalized with respect to $\omega_{m1}$. To avoid the trivial synchronization of two identical magnon modes, we introduce a finite frequency mismatch by choosing $\omega_{m2}=1.01\,\omega_{m1}$ throughout the numerical simulations unless otherwise stated. The remaining parameters are chosen within experimentally accessible ranges reported for WGM-based cavity optomagnonic systems \cite{50,55,56,57} where we have taken $g_1=g_2=2\times10^{-3}\omega_{m1}$, $\kappa_1=\kappa_2=0.010\,\omega_{m1}$, $\gamma_1=\gamma_2=10^{-4}\omega_{m1}$, and $J=0.4\,\omega_{m1}$. To gain insight into the collective magnon dynamics, we first examine the time evolution of the position and momentum quadratures of both the magnon modes before investigating different synchronization regimes which includes complete synchronization, $\phi$-synchronization, and quantum phase synchronization. Following the standard definition adopted in continuous-variable quantum synchronization studies \cite{2,63}, the phase associated with the $j$th magnon mode is obtained from the corresponding semiclassical trajectories as follows

\begin{equation}
\phi_j=
\tan^{-1}
\left(
\frac{p_{js}}
{q_{js}}
\right),
\end{equation}

where $q_{js}$ and $p_{js}$ denote the corresponding steady-state mean values of the magnon quadratures..\

\begin{figure}[p]
\centering

\includegraphics[width=0.80\textwidth]{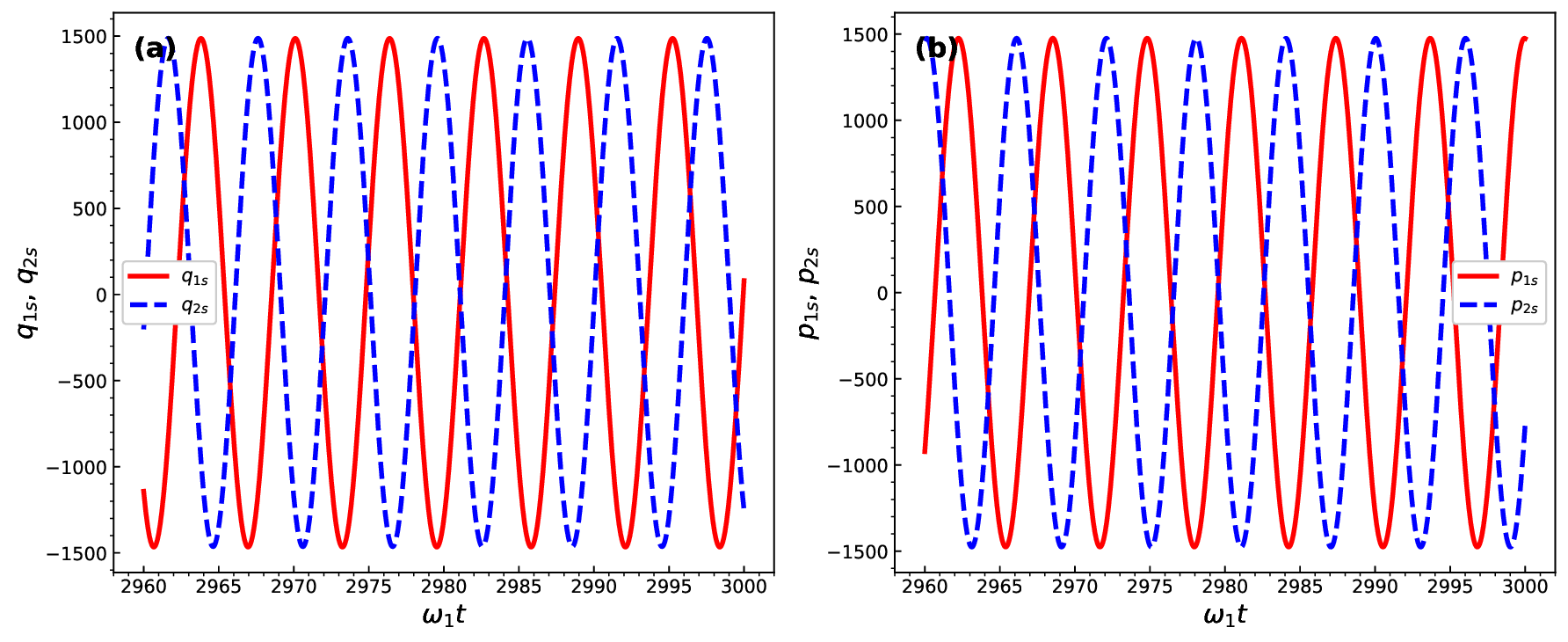}

\vspace{1mm}

\includegraphics[width=0.80\textwidth]{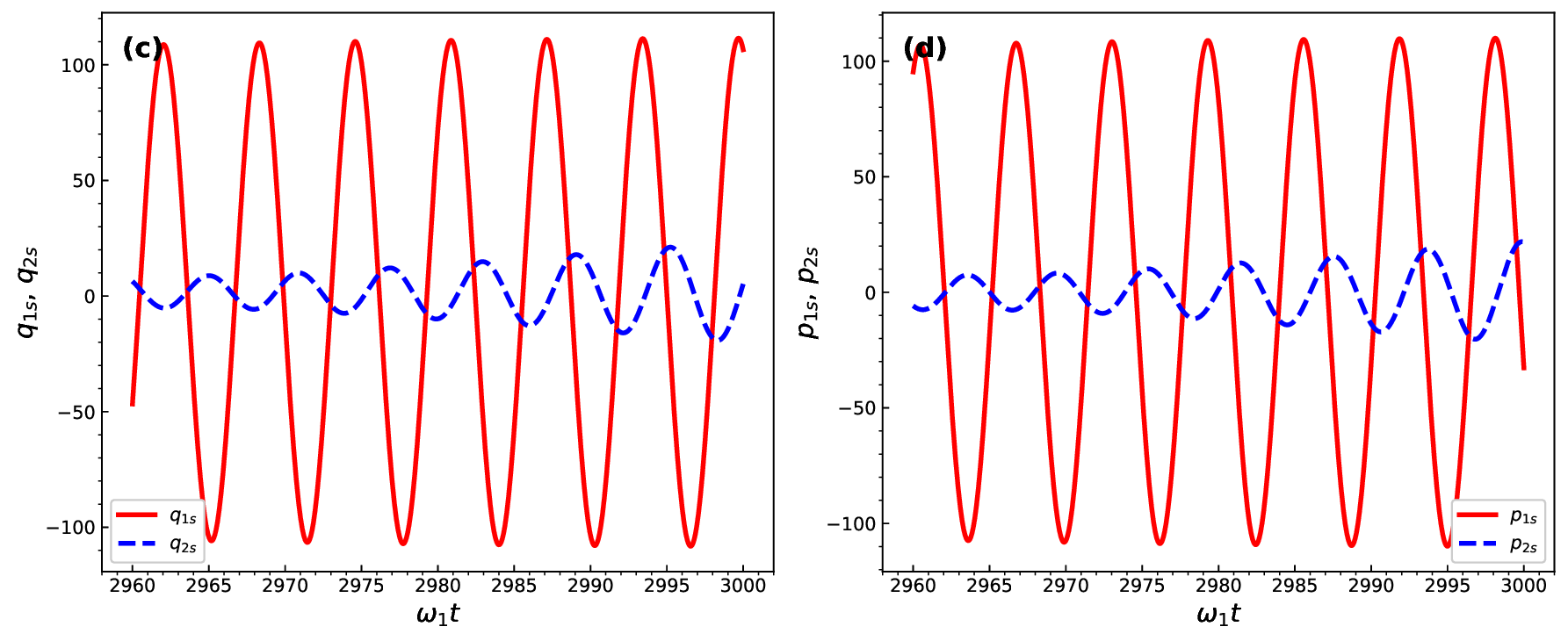}

\vspace{1mm}

\includegraphics[width=0.80\textwidth]{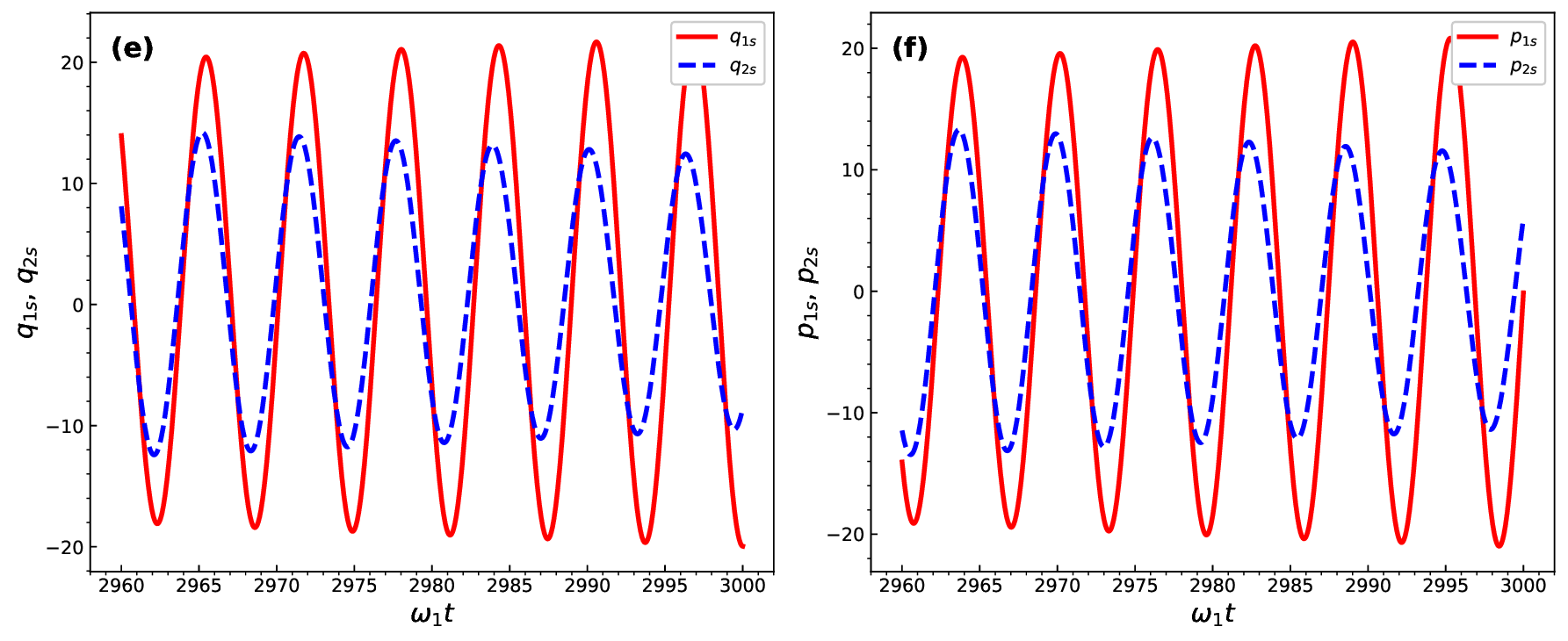}

\vspace{1mm}

\includegraphics[width=0.80\textwidth]{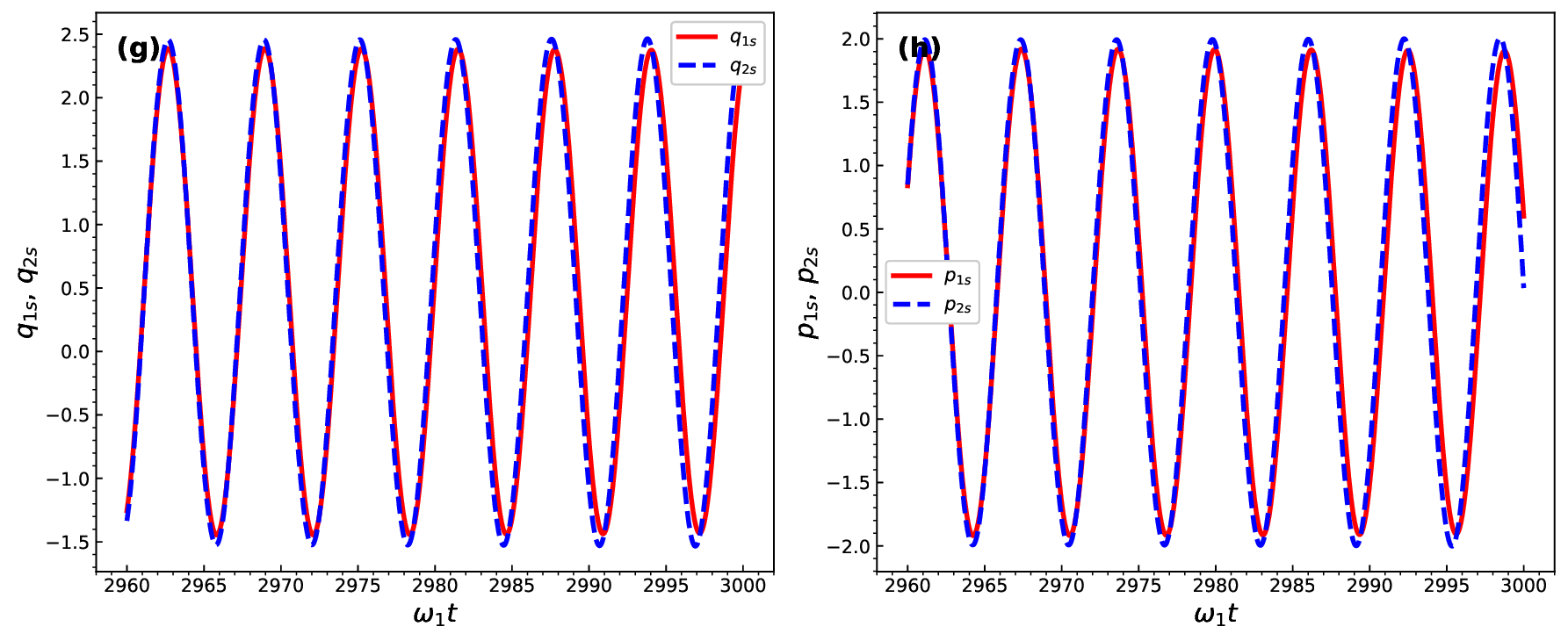}

\caption{Time evolution of the magnon position $q_1$, $q_2$ and momentum  $p_1$, $p_2$ quadratures. For panels (a),(b): $J=0$; for panels (c),(d): $J=0.4\,\omega_{m1}$, $\theta=0$; for panels (e),(f): $J=0.4\,\omega_{m1}$, $\theta=\pi/2$ whereas for panels (g),(h): $J=0.4\,\omega_{m1}$, $\theta=\pi$. Other parameters are given as $\omega_{m2}=1.01\,\omega_{m1}$, $\Delta_a = -1.0\,\omega_{m1}$, $g_{1}=g_{2} = 2\times10^{-3}\,\omega_{m1}$, $\kappa_{1}= \kappa_{2} =  0.010\,\omega_{m1}$, $\gamma_1=\gamma_2=10^{-4}\,\omega_{m1}$, $E_1=E_2=20\,\omega_{m1}$, and $n_{\rm th}=0$.}
\label{fig:mechanical_quadratures_sync}
\end{figure}

We have shown the time evolution of the magnon position and momentum quadratures for different values of the photon-hopping phase factor $\theta$ as shown in Fig.~\ref{fig:mechanical_quadratures_sync}. In the absence of photon hopping, i.e., $J=0$ which are plotted in Figs.~\ref{fig:mechanical_quadratures_sync}(a) and \ref{fig:mechanical_quadratures_sync}(b), the two magnon modes oscillate with comparable amplitudes but exhibit a pronounced phase mismatch. The trajectories remain clearly separated throughout the evolution, indicating that the magnons evolve almost independently without any appreciable synchronization. When the photon-hopping interaction is switched on with $J=0.4,\omega_{m1}$ and $\theta=0$ as shown in Figs.~\ref{fig:mechanical_quadratures_sync}(c) and \ref{fig:mechanical_quadratures_sync}(d), the optical coupling establishes an indirect interaction between the two magnon modes. As a consequence, the dynamics of the second magnon becomes correlated with that of the first magnon. Nevertheless, a significant difference in oscillation amplitudes is still observed, suggesting that the synchronization process remains incomplete. A further increase in the synchronization phenomena is observed for $\theta=\pi/2$ given in Figs.~\ref{fig:mechanical_quadratures_sync}(e) and \ref{fig:mechanical_quadratures_sync}(f). In this case, the amplitude mismatch between the two magnon modes is substantially reduced and the trajectories become noticeably closer.  The oscillations evolve with a much smaller relative deviation which indicates that the phase factor associated with the photon hopping enhances the synchronization dynamics in between the two magnon modes. For $\theta=\pi$ the given Figs.~\ref{fig:mechanical_quadratures_sync}(g) and \ref{fig:mechanical_quadratures_sync}(h) show that the position quadratures $q_1$ and $q_2$ as well as the momentum quadratures $p_1$ and $p_2$ almost overlap during the entire evolution. Both magnons oscillate with nearly identical amplitudes and phases, providing strong evidence of highly synchronized dynamics. The progressive reduction of the mismatch between the two trajectories from $\theta=0$ to $\theta=\pi$ demonstrates that the phase-dependent photon-hopping interaction plays a crucial role in controlling synchronization between the spatially separated magnon modes. 
Overall, the results shown in Fig.~\ref{fig:mechanical_quadratures_sync} show that changing the phase of the photon-hopping interaction provides an effective way to control the collective dynamics of the two magnon modes and hence ultimately drive the proposed system toward a highly synchronized regime.\\

\begin{figure}[t]
\centering
\begin{subfigure}[b]{0.45\textwidth}
\centering
\includegraphics[width=\textwidth]{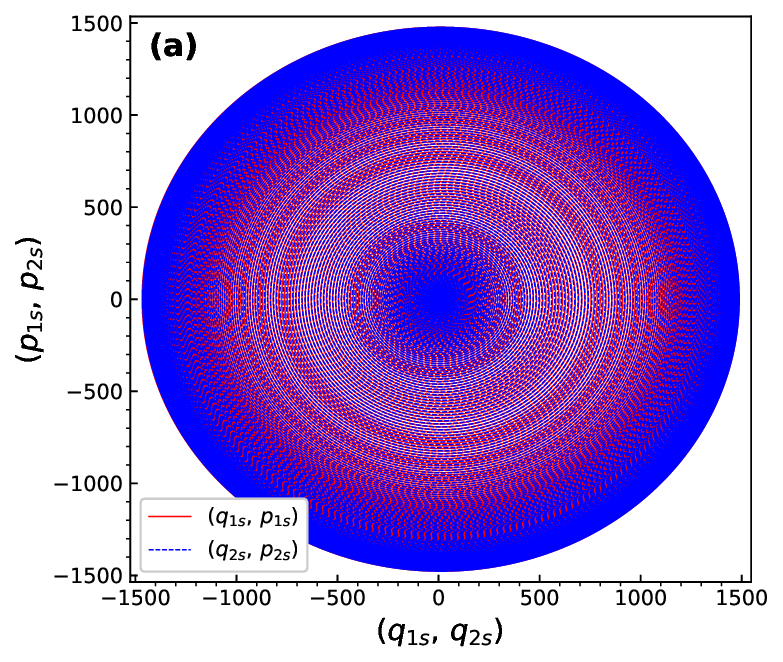}
\end{subfigure}
\begin{subfigure}[b]{0.45\textwidth}
\centering
\includegraphics[width=\textwidth]{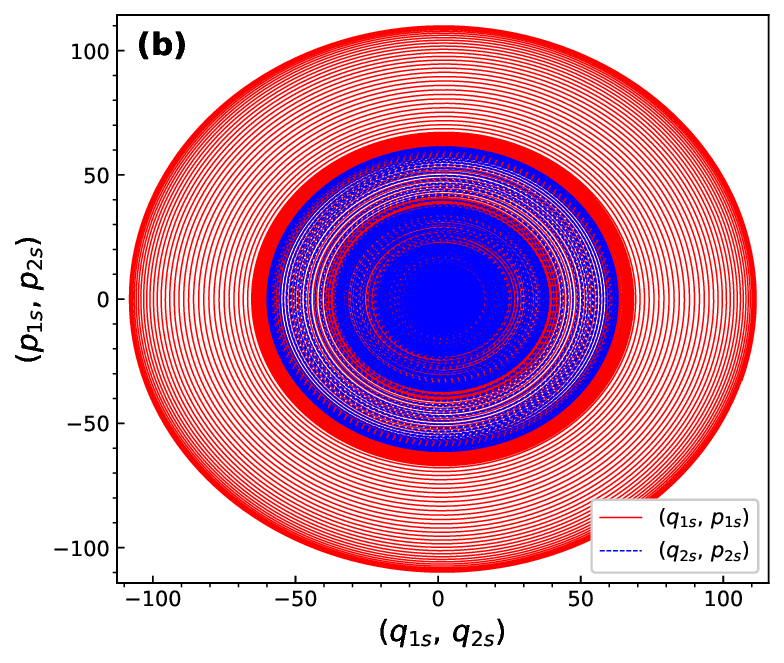}
\end{subfigure}
\begin{subfigure}[b]{0.45\textwidth}
\centering
\includegraphics[width=\textwidth]{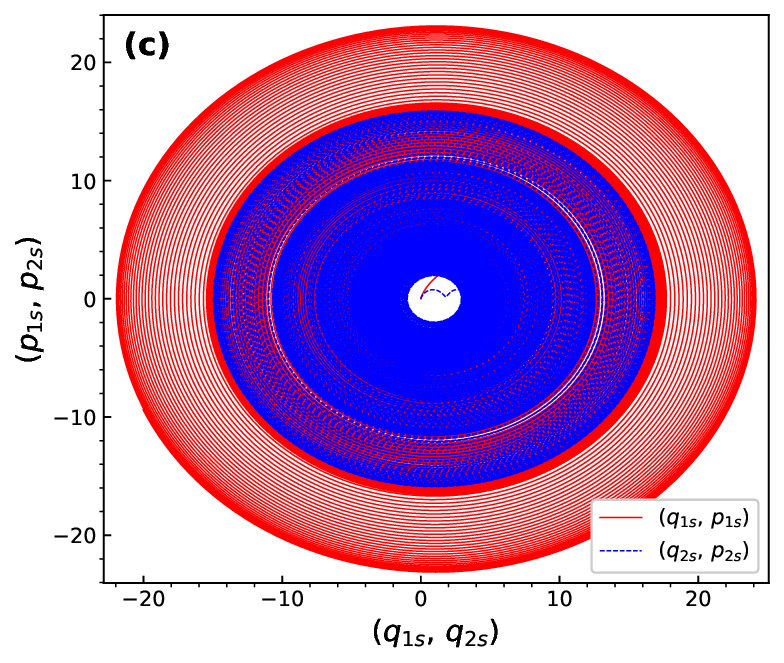}
\end{subfigure}
\begin{subfigure}[b]{0.45\textwidth}
\centering
\includegraphics[width=\textwidth]{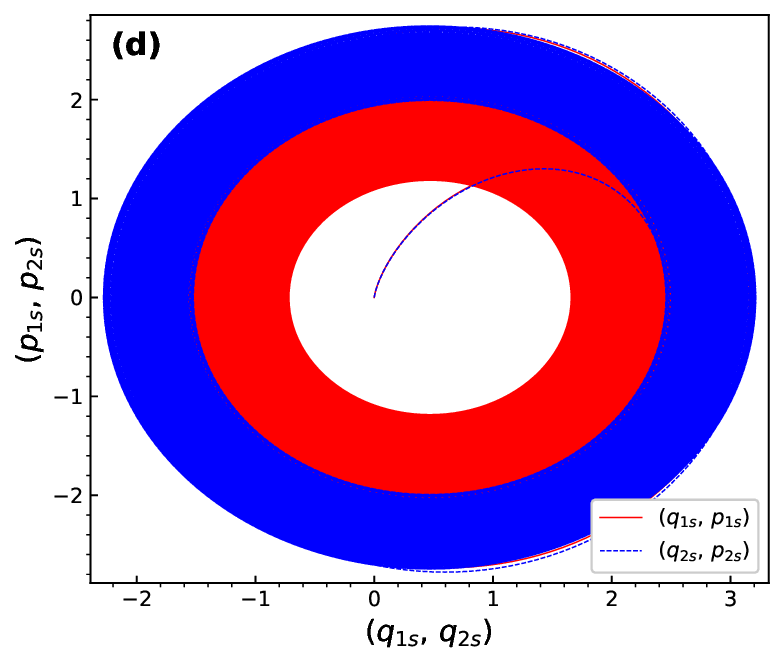}
\end{subfigure}
\caption{
Limit-cycle trajectories of the magnon modes in the $(q_{1s}(t),p_{1s}(t))$ and $(q_{2s}(t),p_{2s}(t))$ phase spaces. For Panel (a): $J=0$; for panel (b): $J=0.4\,\omega_{m1}$, $\theta=0$; for panel (c): $J=0.4\,\omega_{m1}$, $\theta=\pi/2$; whereas for panel (d): $J=0.4\,\omega_{m1}$, $\theta=\pi$. All other parameters are identical to those in Fig.\ref{ fig:mechanical_quadratures_sync}.}
\label{fig:mechanical_limit_cycle_merged}
\end{figure}

To further understand the emergence of synchronization dynamics, we now examine the limit-cycle trajectories of the two magnon modes in phase space, as shown in Fig.~\ref{fig:mechanical_limit_cycle_merged}. In the absence of photon hopping, i.e., $J=0$ which is plotted in Fig.~\ref{fig:mechanical_limit_cycle_merged}(a), the two magnons evolve on distinct large-amplitude limit cycles, indicating that the resonators behave almost independently due to the lack of an interaction channel between them. Consequently, the trajectories remain well separated in phase space and no appreciable synchronization is observed. When the photon-hopping interaction is introduced with $J=0.4,\omega_{m1}$ and $\theta=0$ as given in Fig.~\ref{fig:mechanical_limit_cycle_merged}(b), coherent photon exchange between the optical cavities establishes an indirect coupling between the magnon modes. This photon-mediated interaction reduces the dynamical mismatch between the oscillators, leading to a significant contraction of the limit-cycle area and a noticeable decrease in the separation between the two trajectories. However, the overlap remains incomplete, indicating that the magnons have not yet reached a synchronized state. As the hopping phase is increased to $\theta=\pi/2$  illustrated in Fig.~\ref{fig:mechanical_limit_cycle_merged}(c), the overlap between the two trajectories becomes considerably stronger. Finally, for $\theta=\pi$ which we have shown in Fig.~\ref{fig:mechanical_limit_cycle_merged}(d), the trajectories nearly collapse into a common orbit which demonstrate that both magnons evolve with almost identical amplitudes and phases. The progressive transition from separated limit cycles at $J=0$ to nearly overlapping trajectories at $\theta=\pi$ clearly reveals the crucial role played by the phase-dependent photon-hopping interaction. As the hopping amplitude $J$ provides the indirect coupling pathway through the optical fields, the phase factor $\theta$ governs the effectiveness of this interaction and enables precise control over the collective magnon dynamics and the resulting synchronization behaviour.\\

\begin{figure}[t]
\centering
\begin{subfigure}[b]{0.95\textwidth}
\centering
\includegraphics[width=\textwidth]{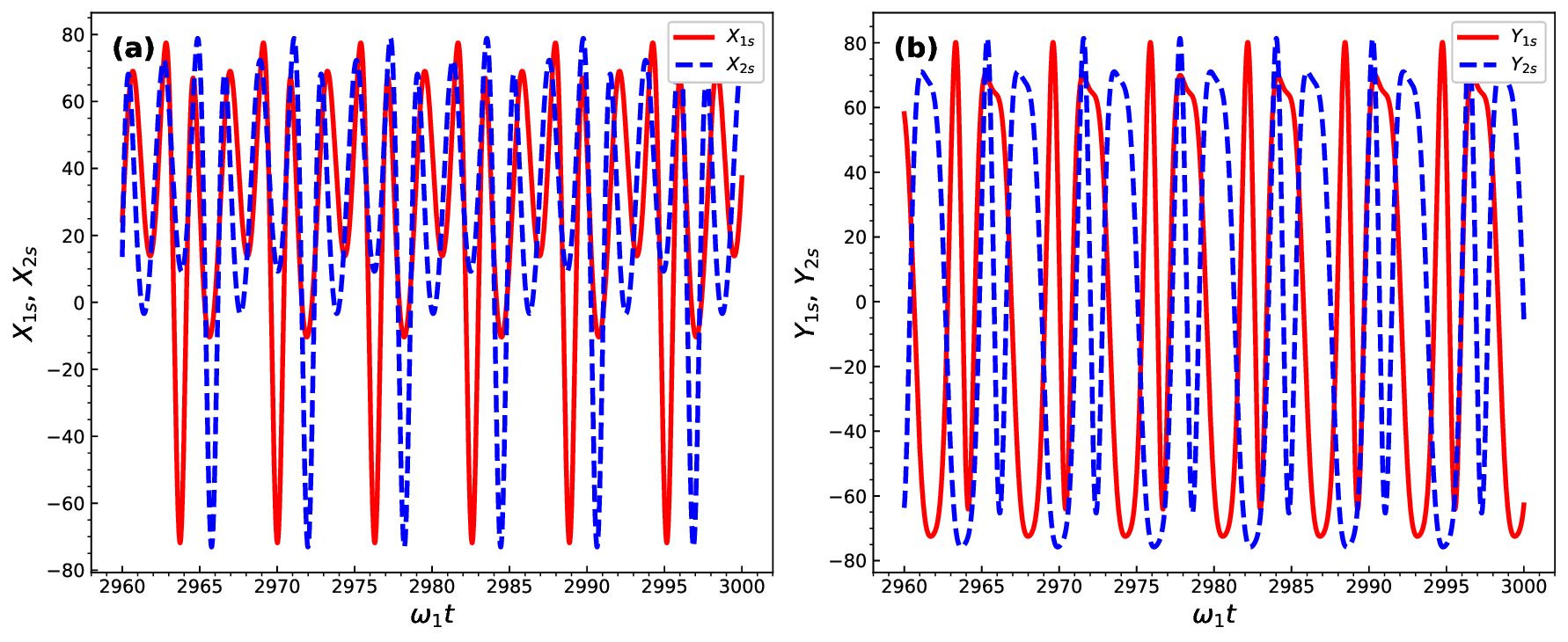}
\end{subfigure}
\begin{subfigure}[b]{0.95\textwidth}
\centering
\includegraphics[width=\textwidth]{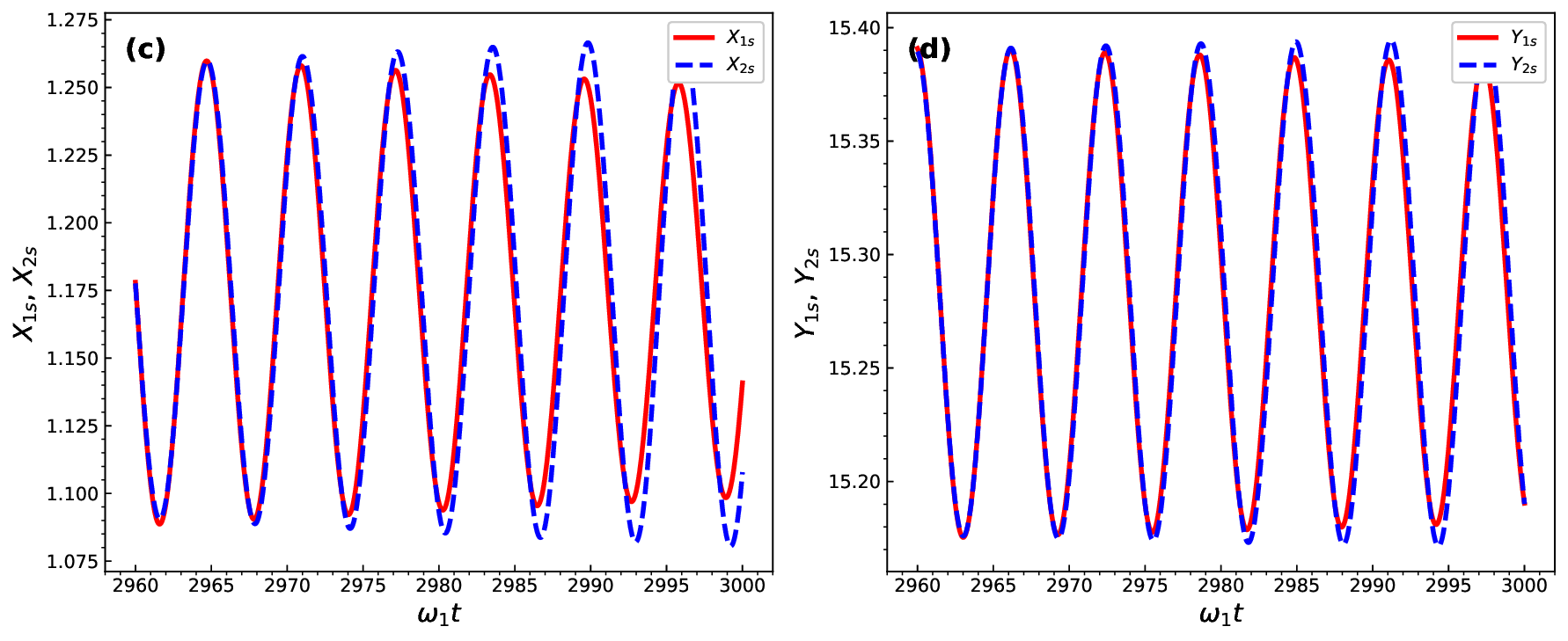}
\end{subfigure}
\caption{Time evolution of the cavity-field quadratures $X_1$ and $X_2$; $Y_1$ and $Y_2$. For panels (a) and (b): $J=0$ whereas for panels (c) and (d): $J=0.4\,\omega_{m1}$, $\theta=\pi$.}
\label{fig:optical_quadratures_sync}
\end{figure}

The synchronization behaviour observed in the magnon subsystem is also reflected in the dynamics of the optical fields. To illustrate this, Fig.~\ref{fig:optical_quadratures_sync} shows the time evolution of the cavity-field quadratures $X_{1s}$, $X_{2s}$ and $Y_{1s}$, $Y_{2s}$ for two representative cases. In the absence of photon hopping, i.e., $J=0$ as shown in Fig.~\ref{fig:optical_quadratures_sync}(a) and Fig.~\ref{fig:optical_quadratures_sync}(b), the optical quadratures associated with the two cavities exhibit distinct oscillatory patterns with noticeable phase and amplitude differences. Since the optical resonators evolve independently in this regime, no efficient exchange of photons occurs between the cavities, resulting in weak dynamical correlations between the corresponding optical fields. Consequently, the cavity modes are unable to effectively mediate synchronization between the two magnon resonators, which is consistent with the unsynchronized dynamics observed in Figs.~\ref{fig:mechanical_quadratures_sync} and \ref{fig:mechanical_limit_cycle_merged}. In contrast, when the photon-hopping interaction is introduced with $J=0.4,\omega_{m1}$ and $\theta=\pi$ as shown in Fig.~\ref{fig:optical_quadratures_sync}(c) and Fig.~\ref{fig:optical_quadratures_sync}(d), the optical quadratures of the two cavities almost completely overlap and oscillate with nearly identical amplitudes and phases. This behavior indicates the establishment of strong coherent correlations between the cavity fields due to efficient photon transport through the hopping channel. As a result, the optical modes provide an effective pathway for transferring dynamical information between the two spatially separated magnon resonators. The strong overlap of the cavity-field trajectories therefore supports the highly synchronized magnon dynamics observed in Figs.~\ref{fig:mechanical_quadratures_sync} and \ref{fig:mechanical_limit_cycle_merged}. These results clearly demonstrate that synchronization in the present system is mediated by the cavity fields and can be efficiently controlled through the phase-dependent photon-hopping interaction.

\begin{figure}[t]
\centering
\begin{subfigure}[b]{0.95\textwidth}
\centering
\includegraphics[width=\textwidth]{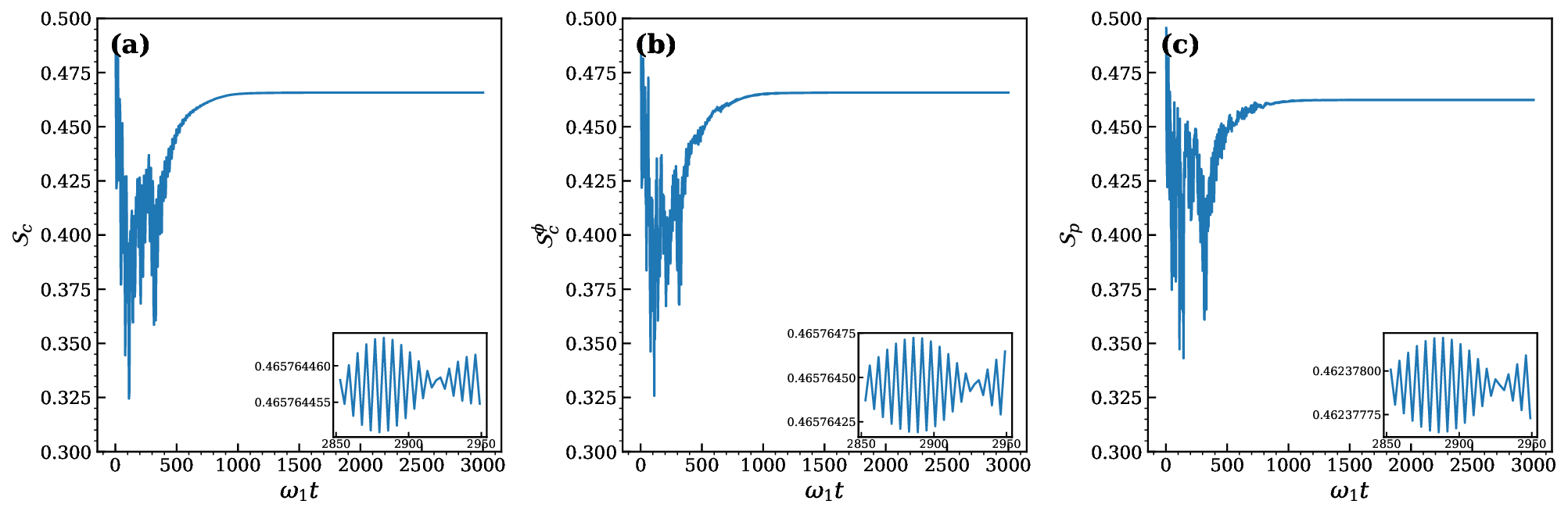}
\end{subfigure}
\begin{subfigure}[b]{0.95\textwidth}
\centering
\includegraphics[width=\textwidth]{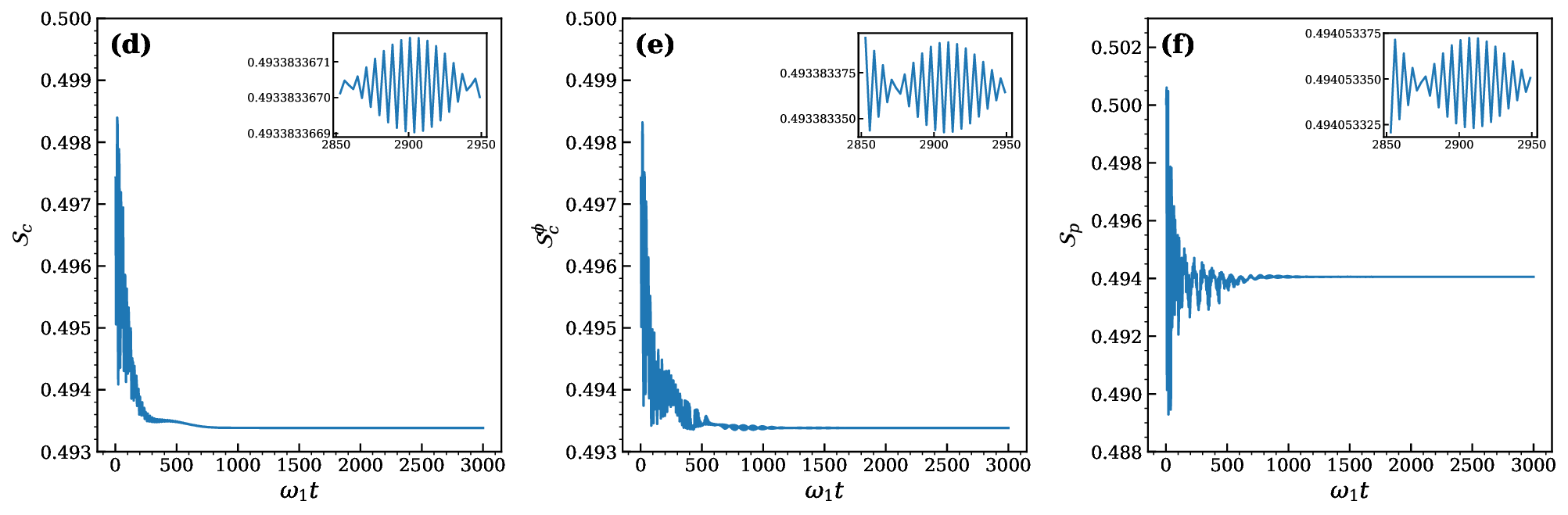}
\end{subfigure}
\begin{subfigure}[b]{0.95\textwidth}
\centering
\includegraphics[width=\textwidth]{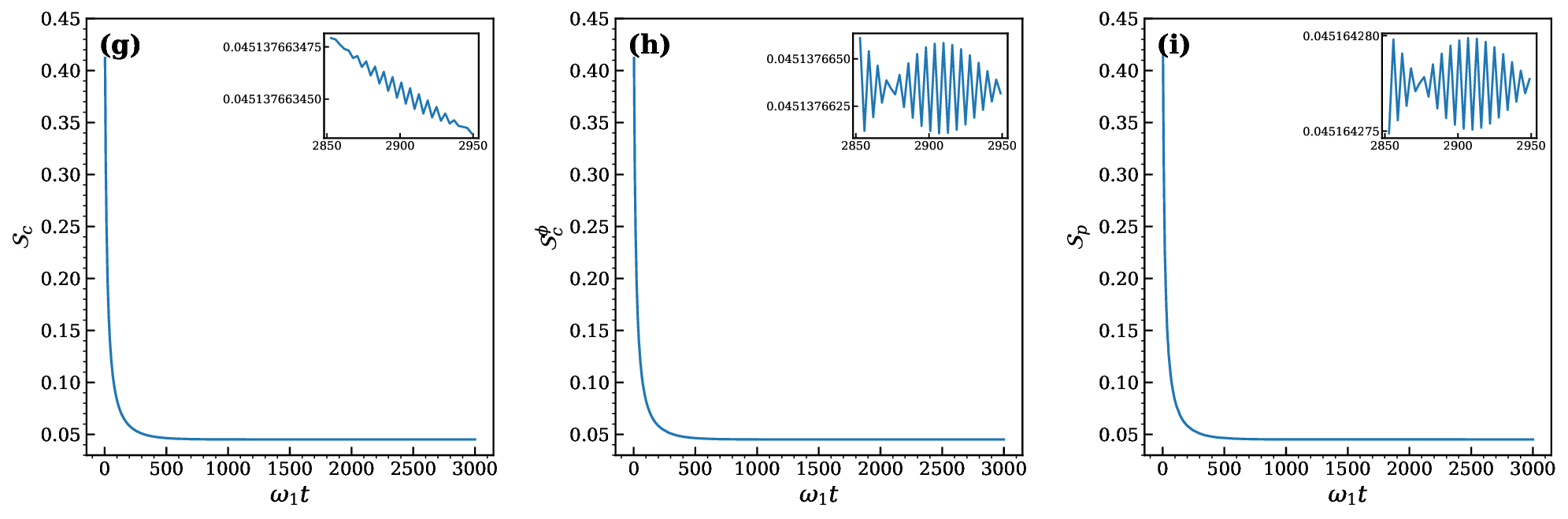}
\end{subfigure}
\caption{Time evolution of the complete quantum synchronization ${S}_{c}$ given in [(a), (d), and (g)]; the phase synchronization given in ${S}_{c}^{\phi}$ [(b), (e), and (h)] and the quantum phase synchronization given in ${S}_{p}$ [(c), (f), and (i)] where panels (a)-(c)represents  $J=0.4\,\omega_{m1}$, $\theta=0$, $n_{\rm th}=0$, panels (d)-(f) for $J=0.4\,\omega_{m1}$, $\theta=\pi$, $n_{\rm th}=0$ whereas  panels (g)-(i) given for $J=0.4\,\omega_{m1}$, $\theta=\pi$, $n_{\rm th}=5$.}
\label{fig:synchronisation_measures_merged}
\end{figure}

The dynamical behaviour observed in Fig.~\ref{fig:mechanical_quadratures_sync}- Fig.~\ref{fig:optical_quadratures_sync} clearly suggests that the phase-dependent photon-hopping interaction plays a crucial role in establishing correlations between the two magnon modes. While the overlap of the trajectories and the corresponding phase-space orbits provides a qualitative indication of synchronization, a more rigorous characterization can be obtained through the synchronization measures introduced in Sec.~II. To this end, Fig.~\ref{fig:synchronisation_measures_merged} shows the time evolution of the complete synchronization measure $S_c$, the $\phi$-synchronization measure $S_c^{\phi}$, and the quantum phase synchronization measure $S_p$ for different values of the hopping phase and thermal occupation number. For $\theta=0$ and $n_{\rm th}=0$ as shown in Fig.~\ref{fig:synchronisation_measures_merged}(a), Fig.~\ref{fig:synchronisation_measures_merged}(b), and Fig.~\ref{fig:synchronisation_measures_merged}(c), all three measures exhibit transient oscillations before gradually approaching steady-state values. The presence of these oscillations reflects the competition between coherent photon-mediated interactions and dissipative processes in the system. Although synchronization is established at long times, the corresponding steady-state values remain relatively moderate, indicating that the photon-hopping channel alone is not sufficient to maximize the correlation between the two magnons. A significant enhancement is observed when the hopping phase is tuned to $\theta=\pi$ which we give in Fig.~\ref{fig:synchronisation_measures_merged}(d), Fig.~\ref{fig:synchronisation_measures_merged}(e), and Fig.~\ref{fig:synchronisation_measures_merged}(f). In this regime, the synchronization measures converge more rapidly and attain larger steady-state values, demonstrating a stronger locking of both amplitudes and phases. This behaviour is fully consistent with the almost complete overlap of the magnons trajectories shown in Fig.~\ref{fig:mechanical_quadratures_sync} and the nearly merged limit cycles shown in Fig.~\ref{fig:mechanical_limit_cycle_merged}. This means the phase factor $\theta$ modifies the interference condition associated with photon transport between the two cavities and thereby strengthens the effective interaction responsible for synchronization. As a consequence, the exchange of dynamical information between the resonators becomes more efficient, leading to a substantial improvement in the synchronization performance. To further investigate the robustness of this synchronized state, we also consider a finite thermal occupation $n_{\rm th}=5$ although keeping $\theta=\pi$ fixed as plotted in Fig.~\ref{fig:synchronisation_measures_merged}(g), Fig.~\ref{fig:synchronisation_measures_merged}(h), and Fig.~\ref{fig:synchronisation_measures_merged}(i). In this case, all synchronization measures are strongly suppressed and rapidly decay to much smaller values. The reduction originates from thermal fluctuations, which introduce additional noise into the magnon dynamics and weaken the coherent correlations established through the cavity fields. Nevertheless, the synchronization measures remain finite throughout the evolution, indicating that the photon-mediated interaction continues to preserve a residual degree of synchronization even in the presence of thermal noise. Overall, Fig.~\ref{fig:synchronisation_measures_merged} quantitatively demonstrates that the  phase factor $\theta$ can be used as an effective control parameter for enhancing synchronization, whereas thermal excitations act against the formation of coherent collective magnon dynamics.

\begin{figure}[t]
\centering
\begin{subfigure}[b]{0.95\textwidth}
\centering
\includegraphics[width=\textwidth]{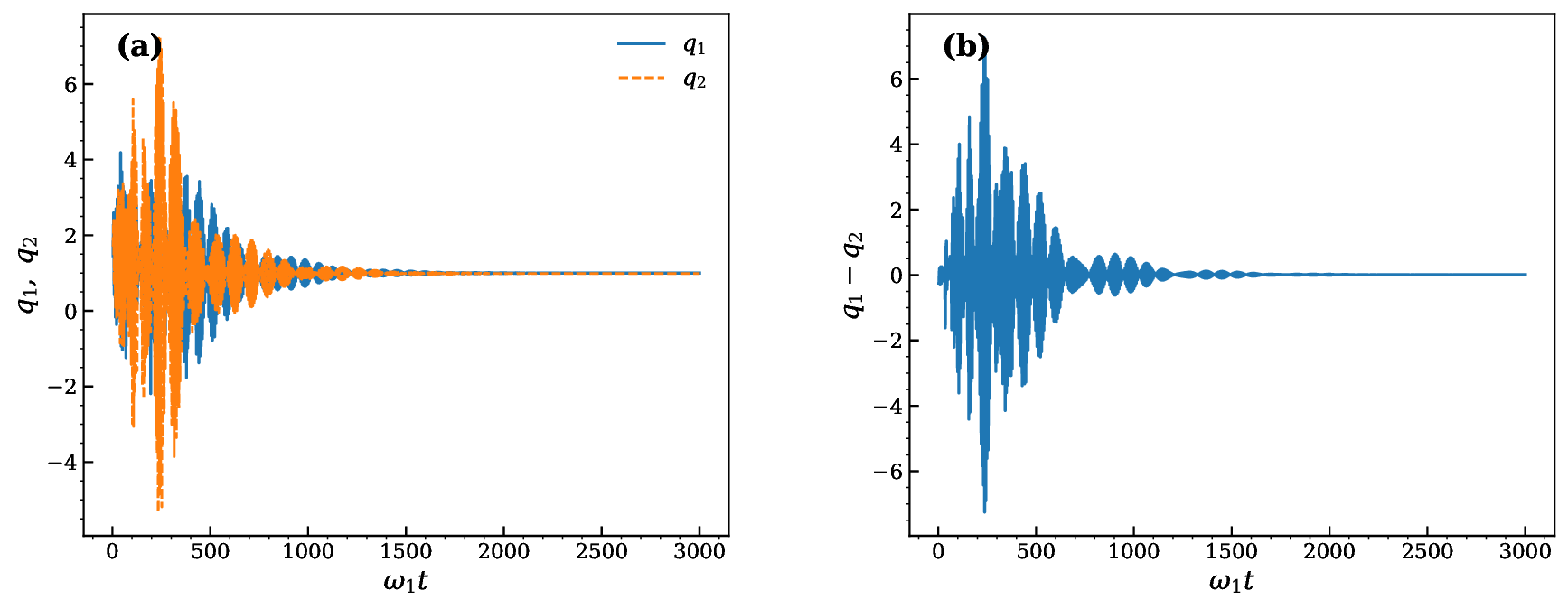}
\end{subfigure}
\begin{subfigure}[b]{0.95\textwidth}
\centering
\includegraphics[width=\textwidth]{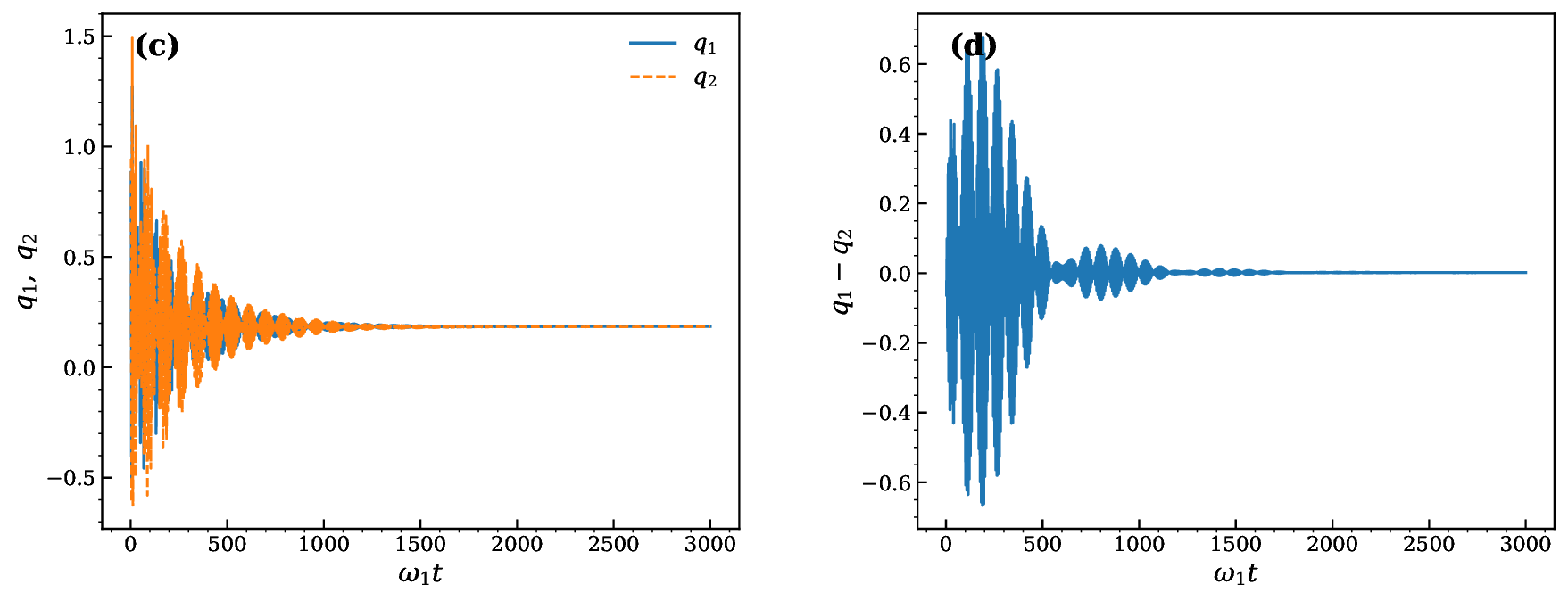}
\end{subfigure}
\caption{Time evolution of the magnon quadratures $q_1$ and $q_2$ together with the synchronization error $q_1-q_2$. Panels (a) and (b) correspond to $\theta=0$ whereas panels (c) and (d) correspond to $\theta=\pi$.}
\label{fig:mechanical_synchronisation_merged}
\end{figure}

To further analyse the synchronization dynamics between the two magnon modes, Fig.~\ref{fig:mechanical_synchronisation_merged} shows the time evolution of the magnon quadratures $q_1$ and $q_2$ together with the corresponding synchronization error $q_1-q_2$ for two different values of the hopping phase. For $\theta=0$  which are plotted in Fig.~\ref{fig:mechanical_synchronisation_merged}(a) and Fig.~\ref{fig:mechanical_synchronisation_merged}(b), the amplitudes of the two magnon modes gradually approach each other after an initial transient period and eventually evolve toward a common steady-state trajectory. As a result, the synchronization error exhibits damped oscillations and progressively decreases with time, indicating the gradual establishment of dynamical correlations between the two magnons. However, the relatively large transient oscillations reveal that the synchronization process remains sensitive to the mismatch between the oscillators during the early stages of the evolution. A markedly different behavior is observed for $\theta=\pi$ which we show  in Fig.~\ref{fig:mechanical_synchronisation_merged}(c) and Fig.~\ref{fig:mechanical_synchronisation_merged}(d). In this case, the two magnon quadratures almost completely overlap throughout the evolution, while the synchronization error is substantially reduced over the entire time domain. The faster suppression of the synchronization error demonstrates that the phase-dependent photon-hopping interaction significantly enhances the locking between the magnon modes. These observations are fully consistent with the synchronization measures presented in Fig.~\ref{fig:synchronisation_measures_merged} where the largest values of $S_c$, $S_c^{\phi}$, and $S_p$ were obtained for $\theta=\pi$. Therefore, Fig.~\ref{fig:mechanical_synchronisation_merged} provides direct evidence that tuning the hopping phase factor $\theta$ not only enhances synchronization but also suppresses the synchronization error, thereby improving the stability and robustness of the synchronized magnon dynamics.

\begin{figure}[t]
\centering
\begin{subfigure}[b]{0.45\textwidth}
\centering
\includegraphics[width=\textwidth]{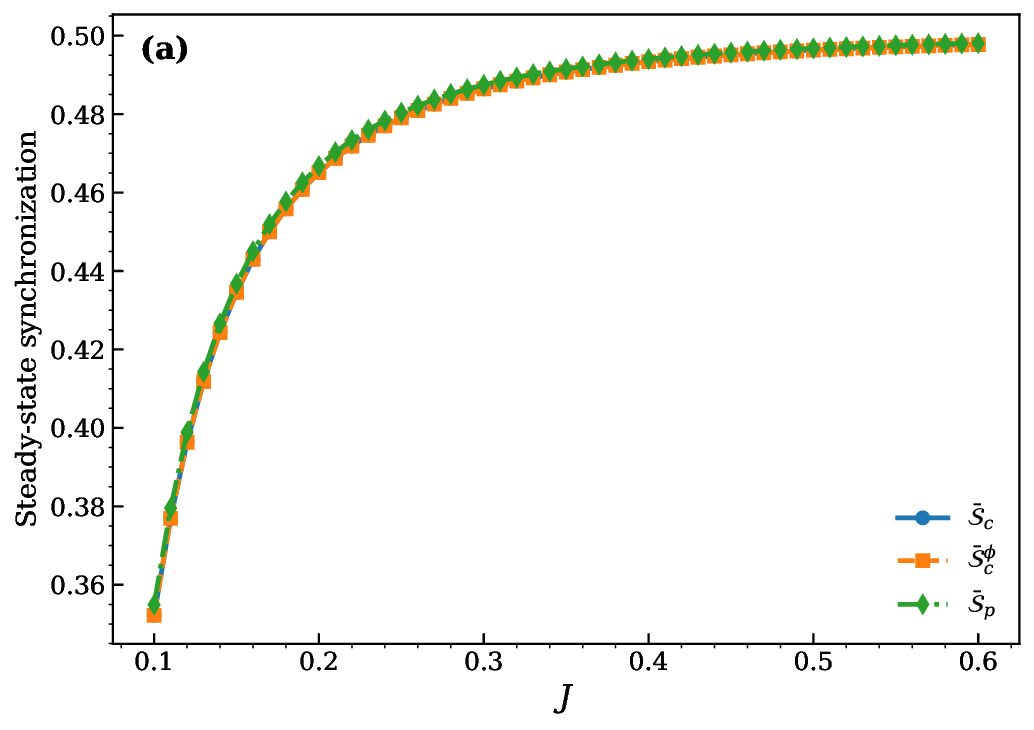}
\end{subfigure}
\begin{subfigure}[b]{0.45\textwidth}
\centering
\includegraphics[width=\textwidth]{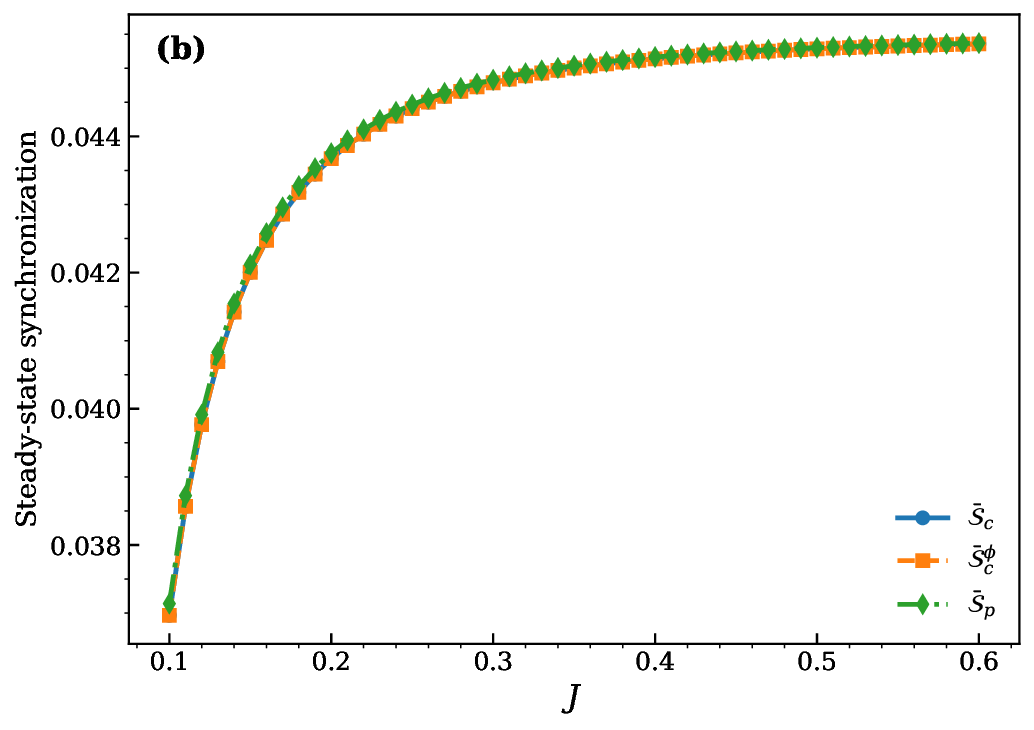}
\end{subfigure}
\caption{Steady-state synchronization $\bar{S}_{c}$, $\bar{S}_{c}^{\phi}$, and $\bar{S}_{p}$ as functions of the photon-hopping strength $J$. Panel (a) corresponds to $n_{\rm th}=0$ whereas for panel (b) $n_{\rm th}=5$.}
\label{fig:Mean_Synchronization_merged}
\end{figure}

To further understand the role of the hopping interaction, Fig.~\ref{fig:Mean_Synchronization_merged} shows the steady-state synchronization measures $\bar{S}_{c}$, $\bar{S}_{c}^{\phi}$, and $\bar{S}_{p}$ as functions of the photon-hopping strength $J$ for two different thermal occupations. As shown in Fig.~\ref{fig:Mean_Synchronization_merged}(a), all three synchronization measures increase monotonically with increasing $J$ when $n_{\rm th}=0$. For weak hopping strengths, the exchange of information between the two cavities remains limited, resulting in a relatively low degree of synchronization. As $J$ increases, coherent photon transport between the cavities becomes more efficient, strengthening the effective interaction between the magnon modes and enhancing the correlation between their amplitudes and phases. Consequently, the synchronization measures increase rapidly before gradually approaching saturation at larger values of $J$. This saturation indicates that strong coherent correlations have already been established and further increasing the hopping strength yields only a minor improvement. The influence of thermal fluctuations is illustrated in Fig.~\ref{fig:Mean_Synchronization_merged}(b), where the thermal occupation is increased to $n_{\rm th}=5$. Although the synchronization measures exhibit a similar dependence on $J$, their magnitudes are noticeably reduced throughout the entire parameter range. This reduction arises from thermal noise, which weakens the coherent correlations generated by the photon-mediated interaction and therefore suppresses synchronization. Nevertheless, all three measures continue to increase with increasing $J$, demonstrating that stronger photon hopping can partially compensate for the detrimental effects of thermal fluctuations. It is also evident that $\bar{S}_{c}$, $\bar{S}_{c}^{\phi}$, and $\bar{S}_{p}$ display nearly identical behaviour over the entire range of hopping strengths, confirming the consistency of the different synchronization criteria employed in this work. These results further establish the photon-hopping strength as an important control parameter for achieving and enhancing synchronization in coupled optomagnonic systems.

\section{Experimental Feasibility}

The proposed scheme can be implemented using currently available cavity optomagnonic platforms based on WGM resonators coupled to YIG spheres. Such systems have been extensively investigated experimentally and provide a suitable architecture for realizing strong interactions between optical and magnon modes through the magneto-optical Faraday effect. In recent experiments, optical cavity decay rates in the range $\kappa/2\pi \sim {\rm MHz}$ and magnon damping rates $\gamma/2\pi \sim {\rm kHz}$ have been routinely achieved, while optomagnonic coupling strengths comparable to those considered in the present work can be obtained under strong optical driving conditions. The magnon frequencies can be conveniently tuned by external bias magnetic fields according to $\omega_{mj}=\gamma H_{Bj}$, allowing precise control of the resonance conditions. The proposed phase-dependent photon hopping can be realized by coupling two WGM resonators through an optical waveguide or optical fiber although phase-controlled schemes have already been demonstrated using optical path engineering and dynamic phase-modulation techniques which indicate that the proposed setup is experimentally feasible. The normalized parameters employed in our simulations, namely $g_j=2\times10^{-3}\omega_{m1}$, $\kappa_j=0.010,\omega_{m1}$, $\gamma_j=10^{-4}\omega_{m1}$, and photon-hopping strengths within the range considered in Fig.~7, fall within experimentally accessible regimes reported in recent cavity optomagnonic and coupled-cavity studies. Therefore, the synchronization effects predicted here, including complete synchronization, $\phi$-synchronization, and quantum phase synchronization, should be observable with existing experimental technology. These considerations indicate that the proposed scheme is experimentally feasible and provides a realistic platform for controlling synchronization in coupled magnonic systems through phase-engineered photon transport.

\section{Conclusion}

In this work, we have theoretically investigated quantum synchronization phenomena in a coupled WGM based cavity optomagnonic system  where each resonator is coupled to a YIG sphere while  the optical WGM cavities are also interconnected through phase-dependent single-photon hopping. We have analysed  complete synchronization, $\phi$-synchronization as well as quantum phase synchronization, and also examined  the effect of single photon  hopping factor on the collective dynamics of the indirectly coupled magnon modes. Our results show that the phase-dependent single photon-hopping interaction plays a key role in establishing synchronization between the two distant magnon modes. As the photon-hopping strength and its phase factor are increased, the magnon trajectories become progressively synchronized and the corresponding limit cycles merge whereas the synchronization error is significantly reduced. These observations are also well supported by the quantum synchronization measures, which consistently predict stronger synchronization for larger hopping phases. We further showed that  gradually increasing the thermal fluctuations significantly suppress it by reducing the coherent correlations between the two magnon modes. The present work demonstrates that the phase-controlled single photon hopping provides an efficient mechanism for manipulating synchronization in the coupled WGM based cavity optomagnonic systems with environmental decoherence and might be useful for studying synchronization-assisted quantum control and coherent information transfer in  magnon-based quantum networks and other hybrid quantum technologies.

\section*{Acknowledgements}
Jia-Xin Peng is supported by the National Natural Science Foundation of China (Grant No.~12504566), the Basic Research Program of Jiangsu (Grant No.~BK20250947), the Natural Science Foundation of the Jiangsu Higher Education Institutions (Grant No.~25KJB140013), and the Natural Science Foundation of Nantong City (Grant No.~JC2024045).


\begin{thebibliography}{99}

\bibitem{1}
C. Huygens,
\textit{Oeuvres Compl\`etes de Christiaan Huygens}
(Martinus Nijhoff, The Hague, 1893), Vol.~15.

\bibitem{2}
A. Mari, A. Farace, N. Didier, V. Giovannetti, and R. Fazio,
``Measures of quantum synchronization in continuous-variable systems,''
Phys. Rev. Lett. \textbf{111}, 103605 (2013).

\bibitem{63}
J. T. Sun, H. D. Liu, and X. X. Yi,
``Quantum synchronization and quantum $\phi$ synchronization in a coupled optomechanical system with Kerr nonlinearity,''
Phys. Rev. A \textbf{109}, 023502 (2024).

\bibitem{3}
L. Ying, Y.-C. Lai, and C. Grebogi,
``Quantum manifestation of a synchronization transition in optomechanical systems,''
Phys. Rev. A \textbf{90}, 053810 (2014).

\bibitem{4}
M. Aspelmeyer, T. J. Kippenberg, and F. Marquardt,
``Cavity optomechanics,''
Rev. Mod. Phys. \textbf{86}, 1391 (2014).

\bibitem{5}
O. V. Zhirov and D. L. Shepelyansky,
``Quantum synchronization and entanglement of two qubits coupled to a driven dissipative resonator,''
Phys. Rev. B \textbf{80}, 014519 (2009).

\bibitem{6}
V. Ameri, M. Eghbali-Arani, A. Mari, A. Farace,
F. Kheirandish, V. Giovannetti, and R. Fazio,
``Mutual information as an order parameter for quantum synchronization,''
Phys. Rev. A \textbf{91}, 012301 (2015).

\bibitem{7}
M. Xu, D. A. Tieri, E. C. Fine, J. K. Thompson, and M. J. Holland,
``Synchronization of two ensembles of atoms,''
Phys. Rev. Lett. \textbf{113}, 154101 (2014).

\bibitem{10}
T. E. Lee, C.-K. Chan, and S. Wang,
``Entanglement tongue and quantum synchronization of disordered oscillators,''
Phys. Rev. E \textbf{89}, 022913 (2014).

\bibitem{13}
M. Samoylova, N. Piovella, G. R. M. Robb, R. Bachelard,
and P. W. Courteille,
``Synchronization of Bloch oscillations by a ring cavity,''
Opt. Express \textbf{23}, 14823 (2015).

\bibitem{14}
Y. G\"ul,
``Synchronization of networked Jahn--Teller systems in SQUIDs,''
Int. J. Mod. Phys. B \textbf{30}, 1650125 (2016).

\bibitem{16}
W. Li, C. Li, and H. Song,
``Quantum synchronization and quantum state sharing in an irregular complex network,''
Phys. Rev. E \textbf{95}, 022204 (2017).

\bibitem{19}
S. Dutta and N. R. Cooper,
``Critical response of a quantum van der Pol oscillator,''
Phys. Rev. Lett. \textbf{123}, 250401 (2019).

\bibitem{22}
J. Zhang, Y.-X. Liu, S. K. \"Ozdemir, R.-B. Wu,
F. Gao, X.-B. Wang, L. Yang, and F. Nori,
``Quantum internet using code-division multiple access,''
Sci. Rep. \textbf{3}, 2211 (2013).

\bibitem{24}
F. Bemani, A. Motazedifard, R. Roknizadeh,
M. H. Naderi, and D. Vitali,
``Synchronization dynamics of two nanomechanical membranes within a Fabry--Perot cavity,''
Phys. Rev. A \textbf{96}, 023805 (2017).

\bibitem{26a}
A. Roulet and C. Bruder,
``Quantum synchronization and entanglement generation,''
Phys. Rev. Lett. \textbf{121}, 063601 (2018).

\bibitem{28a}
G. L. Giorgi, F. Galve, G. Manzano, P. Colet,
and R. Zambrini,
``Quantum correlations and mutual synchronization,''
Phys. Rev. A \textbf{85}, 052101 (2012).

\bibitem{30a}
C.-G. Liao, R.-X. Chen, H. Xie, M.-Y. He,
and X.-M. Lin,
``Quantum synchronization and correlations of two mechanical resonators in a dissipative optomechanical system,''
Phys. Rev. A \textbf{99}, 033818 (2019).

\bibitem{26}
Y. Tabuchi \textit{et al.},
``Hybridizing ferromagnetic magnons and microwave photons in the quantum limit,''
Phys. Rev. Lett. \textbf{113}, 083603 (2014).

\bibitem{27}
X. Zhang, C.-L. Zou, L. Jiang, and H. X. Tang,
``Strongly coupled magnons and cavity microwave photons,''
Phys. Rev. Lett. \textbf{113}, 156401 (2014).

\bibitem{28}
X. Zhang, C.-L. Zou, L. Jiang, and H. X. Tang,
``Cavity magnomechanics,''
Sci. Adv. \textbf{2}, e1501286 (2016).

\bibitem{29}
H. Huebl \textit{et al.},
``High cooperativity in coupled microwave resonator ferrimagnetic insulator hybrids,''
Phys. Rev. Lett. \textbf{111}, 127003 (2013).

\bibitem{30}
C. Kittel,
``On the theory of ferromagnetic resonance absorption,''
Phys. Rev. \textbf{73}, 155--161 (1948).

\bibitem{31}
D. Zhang \textit{et al.},
``Cavity quantum electrodynamics with ferromagnetic magnons in a small yttrium-iron-garnet sphere,''
npj Quantum Inf. \textbf{1}, 15014 (2015).

\bibitem{32}
Q. Cai, J. Liao, B. Shen, G. Guo, and Q. Zhou,
``Microwave quantum illumination via cavity magnonics,''
Phys. Rev. A \textbf{103}, 052419 (2021).

\bibitem{33}
A. Sohail \textit{et al.},
``Controllable Fano-type optical response and four-wave mixing via magnetoelastic coupling in an optomagnomechanical system,''
J. Appl. Phys. \textbf{133}, 154401 (2023).

\bibitem{34}
K. Ullah, M. T. Naseem, and Ö. E. Müstecaplıoğlu,
``Tunable multiwindow magnomechanically induced transparency, Fano resonances, and slow-to-fast light conversion,''
Phys. Rev. A \textbf{102}, 033721 (2020).

\bibitem{36}
J. Li, S.-Y. Zhu, and G. S. Agarwal,
``Magnon-photon-phonon entanglement in cavity magnomechanics,''
Phys. Rev. Lett. \textbf{121}, 203601 (2018).

\bibitem{41}
J. Li and S.-Y. Zhu,
``Entangling two magnon modes via magnetostrictive interaction,''
New J. Phys. \textbf{21}, 085001 (2019).

\bibitem{42}
H. Y. Yuan, S. Zheng, Z. Ficek, Q. Y. He, and M.-H. Yung,
``Enhancement of magnon-magnon entanglement inside a cavity,''
Phys. Rev. B \textbf{101}, 014419 (2020).

\bibitem{44}
G. Zhang, Y. Wang, and J. You,
``Theory of the magnon Kerr effect in cavity magnonics,''
Sci. China Phys. Mech. Astron. \textbf{62}, 987511 (2019).

\bibitem{45}
Z. Zhang, M. O. Scully, and G. S. Agarwal,
``Quantum entanglement between two magnon modes via Kerr nonlinearity driven far from equilibrium,''
Phys. Rev. Res. \textbf{1}, 023021 (2019).

\bibitem{46}
H. Tan and J. Li,
``Einstein--Podolsky--Rosen entanglement and asymmetric steering between distant macroscopic mechanical and magnonic systems,''
Phys. Rev. Res. \textbf{3}, 013192 (2021).

\bibitem{47}
S. V. Kusminskiy, H. X. Tang, and F. Marquardt,
``Coupled spin-light dynamics in cavity optomagnonics,''
Phys. Rev. A \textbf{94}, 033821 (2016).

\bibitem{48}
V. Bittencourt, V. Feulner, and S. V. Kusminskiy,
``Magnon heralding in cavity optomagnonics,''
Phys. Rev. A \textbf{100}, 013810 (2019).

\bibitem{49}
T. Liu, X. Zhang, H. X. Tang, and M. E. Flatté,
``Optomagnonics in magnetic solids,''
Phys. Rev. B \textbf{94}, 060405(R) (2016).

\bibitem{50}
W.-L. Xu, Y.-P. Gao, T.-J. Wang, and C. Wang,
``Magnon-induced optical high-order sideband generation in a hybrid atom-cavity optomagnonic system,''
Opt. Express \textbf{28}, 22334--22344 (2020).

\bibitem{51}
X. Zhang, N. Zhu, C.-L. Zou, and H. X. Tang,
``Optomagnonic whispering-gallery microresonators,''
Phys. Rev. Lett. \textbf{117}, 123605 (2016).

\bibitem{52}
A. Osada \textit{et al.},
``Cavity optomagnonics with spin-orbit coupled photons,''
Phys. Rev. Lett. \textbf{116}, 223601 (2016).

\bibitem{53}
J. A. Haigh, A. Nunnenkamp, A. J. Ramsay, and A. J. Ferguson,
``Triple-resonant Brillouin light scattering in magneto-optical cavities,''
Phys. Rev. Lett. \textbf{117}, 133602 (2016).

\bibitem{54}
J. Graf, H. Pfeifer, F. Marquardt, and S. V. Kusminskiy,
``Cavity optomagnonics with magnetic textures: Coupling a magnetic vortex to light,''
Phys. Rev. B \textbf{98}, 241406(R) (2018).

\bibitem{55}
K. Wang, Y.-P. Gao, R. Jiao, and C. Wang,
``Recent progress on optomagnetic coupling and optical manipulation based on cavity optomagnonics,''
Front. Phys. \textbf{17}, 42201 (2022).

\bibitem{56}
X. Pan, S.-P. Liu, T. Shui, and W.-X. Yang,
``Optical quadrature squeezing via the Faraday effect in cavity optomagnonics,''
J. Opt. Soc. Am. B \textbf{40}, 3065--3072 (2023).

\bibitem{57}
F. H. A. Mathkoor, S. K. Singh, R. Ahmed, \textit{et al.},
``Bipartite entanglement and Gaussian quantum steering in a whispering gallery mode coupled with two magnon modes,''
Sci. Rep. \textbf{15}, 13503 (2025).

\bibitem{60}
T. Holstein and H. Primakoff,
``Field dependence of the intrinsic domain magnetization of a ferromagnet,''
Phys. Rev. \textbf{58}, 1098--1113 (1940).

\bibitem{61}
C. W. Gardiner and P. Zoller,
\textit{Quantum Noise: A Handbook of Markovian and Non-Markovian Quantum Stochastic Methods with Applications to Quantum Optics}
(Springer, Berlin, 2004).

\bibitem{62}
C. Kittel,
``On the theory of ferromagnetic resonance absorption,''
Phys. Rev. \textbf{73}, 155--161 (1948).

\end{thebibliography}
\end{document}